\documentclass[10pt, letterpaper, onecolumn]{emulateapj}

\def\Msun{M_\odot}
\def\microas{\mu{\rm as}}

\begin{document}

\title{Masses, Luminosities, and Orbital Coplanarities of the $\mu$ Orionis 
Quadruple Star System from PHASES Differential Astrometry}

\author{Matthew W.~Muterspaugh\altaffilmark{1}, 
Benjamin F.~Lane\altaffilmark{2}, Francis C.~Fekel\altaffilmark{3}, 
Maciej Konacki\altaffilmark{4}, Bernard F.~Burke\altaffilmark{5}, 
S.~R.~Kulkarni\altaffilmark{6}, M.~M.~Colavita\altaffilmark{7}, 
M.~Shao\altaffilmark{7}, Sloane J.~Wiktorowicz\altaffilmark{8}}
\altaffiltext{1}{Townes Fellow, University of California, 
Space Sciences Laboratory, 7 Gauss Way, Berkeley, CA 94720-7450}
\altaffiltext{2}{Draper Laboratory,  555 Technology Square, Cambridge, MA 
02139-3563}
\altaffiltext{3}{Tennessee State University, Center of Excellence in 
Information Systems, 3500 John A. Merritt Blvd., Box No.~9501, Nashville, TN 
37203-3401}
\altaffiltext{4}{Nicolaus Copernicus Astronomical Center, Polish Academy of 
Sciences, Rabianska 8, 87-100 Torun, Poland}
\altaffiltext{5}{MIT Kavli Institute for Astrophysics and Space Research, 
MIT Department of Physics, 70 Vassar Street, Cambridge, MA 02139}
\altaffiltext{6}{Division of Physics, Mathematics and Astronomy, 105-24, 
California Institute of Technology, Pasadena, CA 91125}
\altaffiltext{7}{Jet Propulsion Laboratory, California Institute of 
Technology, 4800 Oak Grove Dr., Pasadena, CA 91109}
\altaffiltext{8}{Department of Geological and Planetary Sciences, California 
Institute of Technology, Pasadena, CA 91125}

\email{matthew1@ssl.berkeley.edu, blane@mit.edu, maciej@ncac.torun.pl}

\begin{abstract}
$\mu$ Orionis was identified by spectroscopic studies as a quadruple star 
system.  Seventeen high precision differential astrometry measurements of 
$\mu$ Ori have been collected by the Palomar High-precision Astrometric Search 
for Exoplanet Systems (PHASES).  These show both the motion of the long 
period binary orbit and short period perturbations superimposed on that 
caused by each of the components in the long period system being themselves 
binaries.  The new measurements enable the orientations of the long period 
binary and short period subsystems to be determined.  Recent theoretical 
work predicts the
distribution of relative inclinations between inner and outer orbits of 
hierarchical systems to peak near 40 and 140 degrees.  
The degree of coplanarity of this 
complex system is determined, and the angle between the planes of the A-B
and Aa-Ab orbits is found to be $136.7 \pm 8.3$ degrees, near the predicted
distribution peak at 140 degrees; this result is discussed in the context of 
the handful of systems with established mutual inclinations.  The system
distance and masses for each component are obtained from a combined fit
of the PHASES astrometry and archival radial velocity observations. 
The component masses have
relative precisions of $5\%$ (component Aa), $15\%$ (Ab), and 
$1.4\%$ (each of Ba and Bb).  The median size of the minor axes of the 
uncertainty ellipses for the new measurements is 20 micro-arcseconds 
($\microas$).  Updated orbits for $\delta$ Equulei, $\kappa$ Pegasi, and V819 
Herculis are also presented.
\end{abstract}


\section{Introduction}

$\mu$ Orionis (61 Ori, HR 2124, HIP 28614, HD 40932) is a quadruple star 
system that has been extensively studied by radial velocity (RV) and 
differential astrometry.  It is located just North of Betelgeuse, Orion's 
right shoulder (left on the sky); $\mu$ Ori is a bright star that is visible 
to the unaided eye even in moderately light-polluted skies.  \cite{Frost1906} 
discovered it to be a short period (four and a half day) single-lined 
spectroscopic binary; this was component Aa, whose short-period, low mass 
companion Ab has never been detected directly.  \cite{Aitken1914} discovered 
it also had a more distant component (B) forming a sub-arcsecond visual 
binary.  Much later, \cite{Fekel1980} found B was itself a short-period 
(4.78 days) double-lined spectroscopic binary, making the system quadruple; 
these stars are designated Ba and Bb.  Most recently, \cite{Fekel2002} 
(hereafter F2002) reported the astrometric orbit of the A-B motion, 
double-lined RV orbits for A-B and the Ba-Bb subsystem, and a single-lined RV 
orbit for the Aa-Ab subsystem.  F2002 estimate the spectral types as A5V (Aa, 
an Am star), F5V (Ba), and F5V (Bb), though they note these are 
classifications are less certain than usual due to the complexity of the 
system.  For a more complete discussion of the history of $\mu$ Ori, see F2002.

Until now, astrometric observations have only been able to characterize the 
long period A-B motion, lacking the precision necessary to measure the 
astrometric perturbations to this orbit caused by the Aa-Ab and Ba-Bb 
subsystems.  The method described by \cite{LaneMute2004a} for ground-based 
differential astrometry at the $\sim 20$ $\microas$ level for sub-arcsecond 
(``speckle'') binaries has been used to study $\mu$ Ori during the 2004-2007 
observing seasons.  These measurements represent an improvement in precision 
of over two orders of magnitude over previous work on this system.

The goal of the current investigation is to report the center-of-light 
(photocenter) astrometric orbits of the Aa-Ab and Ba-Bb subsystems.  This 
enables measurement of the coplanarities of the A-B, Aa-Ab, and Ba-Bb 
orbits.  The masses and luminosity ratio of Aa and Ab are measured for the 
first time.Also presented are updated orbits for the PHASES targets 
$\delta$ Equ, $\kappa$ Peg, and V819 Her.

Astrometric measurements were made at the Palomar Testbed Interferometer 
\citep[PTI;][]{col99} as part of the Palomar High-precision Astrometric 
Search for Exoplanet Systems (PHASES) program \citep{Mute06Limits}.  PTI is 
located on Palomar Mountain near San Diego, CA. It was developed by the Jet 
Propulsion Laboratory, California Institute of Technology for NASA, as a 
testbed for interferometric techniques applicable to the Keck Interferometer 
and other missions such as the Space Interferometry Mission (SIM).  It 
operates in the J ($1.2 \mu{\rm m}$), H ($1.6 \mu{\rm m}$), and K 
($2.2 \mu{\rm m}$) bands, and combines starlight from two out of three 
available 40-cm apertures. The apertures form a triangle with one 110 and two 
87 meter baselines.

\section{Observations and Data Processing}

\subsection{PHASES Observations}

\subsubsection{Instrumental Setup}

$\mu$ Ori was observed with PTI on 17 nights in 2004-2007 with the observing 
mode described in \cite{LaneMute2004a}.  Starlight is collected from 
two apertures, collimated, and sent to a central beam combining facility.  
There, the light from each telescope reflects from movable mirrors (delay 
lines) whose position is constantly varied to account for sidereal motion 
and to track atmospheric index of refraction variations.  After this first 
set of delay lines, a beam-splitter is used to divide the light from each 
telescope; $\sim 70\%$ of the light is sent to an interferometric beam 
combiner that monitors a single fringe from any star in the field at rapid 
(10-20 millisecond) time-scales to measure fringe phase variations caused by 
the atmosphere, and provide feedback to the main delay lines.  This 
process phase-stabilizes the other $\sim 30\%$ of the light 
\citep[a technique known as phase referencing, ][]{lc03}, which is sent to a 
second interferometric beam combiner that can add an additional variable 
delay of order $250\,{\rm \mu m}$ to the light from one telescope.  This 
variable delay is modulated with a triangle waveform, scanning through 
interferograms from all stars within the subarcsecond field of view.  These 
interferogram scans are the observables used for PHASES astrometry.

\subsubsection{Data Reduction}

Modifications to the data processing algorithm since the original report are 
given by \cite{Mut05_delequ} and have been incorporated in the current 
study.  Interferogram templates are fit to each scan, forming a likelihood 
function of the separations of interferograms formed by components A and B.  A 
grid of differential right ascension and declination is formed, and a $\chi^2$ 
likelihood surface is mapped onto this grid by converting delay separation to 
differential right ascension and declination.  That $\chi^2$ surface is 
coadded over all the scans of $\mu$ Ori within the night (typically 
$\sim 1000$ scans or more).  The deepest minimum in the $\chi^2$ surface 
corresponds to the binary separation, while the width of that minimum 
determines the uncertainty ellipse.  Due to the oscillatory nature of the 
interferograms, other local minima can exist; these ``sidelobes'' are 
separated by the interferometer's resolution 
$\sim \lambda / B \sim 4\,{\rm mas}$ ($B$ is the separation between the 
telescopes, and $\lambda$ the wavelength of starlight), an amount much larger 
than the width of an individual minimum.  The SNR can be increased by coadding 
many scans and by earth-rotation synthesis, which smears all but the true 
minimum, and the true minimum then can be established.  Only those 
measurements for which no sidelobes appear at the $4\sigma$ contour of the 
deepest minimum are used in orbit fitting.  

All measurements have been processed using this new data reduction pipeline.  
The measurements are listed in Table \ref{phasesDataMuOri}, in the 
ICRS 2000.0 reference frame.

\subsubsection{Technique Upgrades}

Data since mid-2006 have benefited from the use of an automatic alignment 
system and longitudinal dispersion compensator; the affected data points are 
noted in Table \ref{phasesDataMuOri}.  These modifications reduced the 
throughput of the astrometry setup; to compensate, a 50 Hz phase tracking rate 
was sometimes used, whereas observations previous to these changes utilized 
100 Hz tracking for monitoring the atmosphere.

Drifts in optical alignment may result in variable pupil sampling at the 
interferometer apertures, changing the effective interferometric baseline.  To 
minimize this potential systematic, a continuous realignment system has been 
developed.  A red laser is coaligned with the starlight and reverse 
propagated through the interferometer.  Four-percent reflective pellicle 
beamsplitters are placed near the focuses of the interferometer telescopes to 
extract this tracer beam and redirect to a camera where the pupil is reimaged.
The angles of the first flat mirrors receiving incoming starlight in the 
beam combining lab are continually adjusted by a closed-loop feedback system 
to hold the laser spot on the camera at the telescope.

The path compensation for the geometric delay at PTI has been 
done with movable mirrors in air, which has a wavelength-dependent index 
of refraction.  The fringe packets of astrophysical sources are 
dispersed by an amount that depends on the difference in air paths 
between arms of the interferometer; this changes the shape and overall 
location of the fringe packets.  To compensate, two prisms are introduced in 
each of the interferometer's two arms.  The set in one arm is static.  The 
other pair are slid relative to each other along the slope of the prism to 
introduce a variable amount of glass dispersion whose shape is opposite that 
of air to high order.  This flattens the variability of delay versus 
wavelength.  The system is calibrated by setting the telescope siderostats 
into a retroreflecting mode, using an internal white-light source to form 
interferograms, which are detected with a low-resolution spectrometer (5 
elements across K band), and measuring the offsets between the 
interferograms as a function of prism location.  During observations, the 
prism position is continuously changed with an open loop control calculated 
from the locations of the delay lines.

There is insufficient new data to establish the degree two which these 
instrumental changes might be reducing excess data scatter or to establish a 
relative weighting between data subsets.  No large discontinuities in the 
orbital motions are seen between pre- and post-upgrade subsets for $\mu$ Ori 
or the other PHASES targets, suggesting the subsets are compatible.  For 
the purposes of the current investigation, the PHASES observations are 
treated as a single data set with equal weighting on observations from before 
and after these upgrades.

\subsubsection{The PHASES Astrometric Orbits}

The differential astrometry measurements are listed in Table 
\ref{phasesDataMuOri}, in the ICRS 2000.0 reference frame.  The existence of 
data scatter beyond the level estimated by formal uncertainties from the 
PHASES analysis algorithm was determined by model fitting the PHASES data 
alone.  Model fitting was performed with standard $\chi^2$ sum of squared 
residuals minimization, slightly complicated by the two dimensional 
nature of the uncertainty ellipses but still straight-forward to carry out.  
The limited number and time span of the PHASES observations prevents 
an independent fitting of that data set to a 4-body, 3-Keplerian model to 
determine potential noise excess.  Thus the A-B period and Aa-Ab and Ba-Bb 
periods, eccentricities, and angles of periastron passages were fixed at the 
values reported in F2002 (in the case of Ba-Bb, which had zero eccentricity 
in F2002, the angle of periastron passage is undefined and fixed at zero).  
The minimum $\chi^2$ does not equal the 
number of degrees of freedom.  An excess noise factor of 1.73 is found, and 
the PHASES uncertainties reported in Table \ref{phasesDataMuOri} have been 
increased by this amount over the formal uncertainties.  The rescaled (raw) 
median minor- and major-axis uncertainties are 20 (11) and 
347 (200) $\microas$.  The rescaled (raw) mean minor- and 
major-axis uncertainties are 27 (16) and 668 (386) $\microas$.


\begin{deluxetable}{lllllllllllll}
\tabletypesize{\scriptsize}
\tablecolumns{13}
\tablewidth{0pc} 
\tablecaption{PHASES data for $\mu$ Ori\label{phasesDataMuOri}}
\tablehead{ 
\colhead{HJD-2400000.5} & \colhead{$\delta$RA}    & \colhead{$\delta$Dec}  & \colhead{$\sigma_{\rm min}$} & 
\colhead{$\sigma_{\rm maj}$} & \colhead{$\phi_{\rm e}$} &  \colhead{$\sigma_{\rm RA}$} & \colhead{$\sigma_{\rm Dec}$} 
& \colhead{$\frac{\sigma_{\rm RA, Dec}^2}{\sigma_{\rm RA}\sigma_{\rm Dec}}$} & \colhead{N} & 
\colhead{LDC} & \colhead{Align} & \colhead{Rate} \\
\colhead{}             &  \colhead{(mas)}        & \colhead{(mas)}        & \colhead{($\microas$)}       & 
\colhead{($\microas$)}       & \colhead{(deg)}          & \colhead{($\microas$)}       & \colhead{($\microas$)}     
& \colhead{}                                                                 & \colhead{}  & 
\colhead{}    & \colhead{}      &  \colhead{(Hz)}
}
\startdata
53271.49964 & 59.1469 & 105.9933 & 9.1 & 487.1 & 146.70 & 407.2 & 267.6 
& -0.99918 & 3092 & 0 & 0 & 100 \\
53285.47060 & 61.8963 & 110.1620 & 14.6 & 376.9 & 19.95 & 354.3 & 129.4 
& 0.99273 & 2503  & 0 & 0 & 100 \\
53290.47919 & 62.4497 & 112.0044 & 39.6 & 2034.0 & 151.07 & 1780.2 & 984.6 
& -0.99894 & 531  & 0 & 0 & 100 \\
53312.46161 & 66.0452 & 119.6805 & 5.2 & 108.2 & 158.79 & 100.9 & 39.5 
& -0.98982 & 6840 & 0 & 0 & 100 \\
53334.41210 & 69.4569 & 127.2488 & 11.5 & 235.3 & 160.77 & 222.2 & 78.3 
& -0.98775 & 2876 & 0 & 0 & 100 \\
53340.37952 & 70.2141 & 128.6951 & 15.0 & 165.2 & 157.64 & 152.9 & 64.4 
& -0.96791 & 2056 & 0 & 0 & 100 \\
53341.34723 & 70.4709 & 129.3402 & 13.2 & 146.6 & 153.14 & 130.9 & 67.3 
& -0.97539 & 3570 & 0 & 0 & 100 \\
53639.51295 & 104.4652 & 211.4488 & 16.3 & 226.6 & 149.61 & 195.6 & 115.5 
& -0.98653 & 1204 & 0 & 0 & 100 \\
53663.47636 & 106.2444 & 217.1615 & 10.4 & 121.4 & 153.87 & 109.0 & 54.3 
& -0.97702 & 1829 & 0 & 0 & 100 \\
53698.43902 & 110.2454 & 225.7804 & 46.8 & 894.1 & 37.76 & 707.5 & 548.8 
& 0.99417 & 827   & 0 & 0 & 100 \\
53705.34654 & 108.6092 & 226.3868 & 56.3 & 1797.7 & 151.29 & 1576.9 & 864.9 
& -0.99724 & 574  & 0 & 0 & 100 \\
53732.29696 & 111.8048 & 231.4902 & 20.0 & 187.4 & 156.29 & 171.8 & 77.5 
& -0.95961 & 1103 & 0 & 0 & 100 \\
53753.23270 & 113.4891 & 235.7197 & 43.7 & 346.8 & 158.08 & 322.2 & 135.7 
& -0.93783 & 621  & 0 & 0 & 100 \\
53789.18100 & 117.7579 & 244.2916 & 47.3 & 2294.6 & 36.86 & 1836.0 & 1377.0 
& 0.99908 & 610   & 0 & 0 & 100 \\
54056.41922 & 131.6278 & 287.8511 & 35.5 & 279.5 & 159.04 & 261.3 & 105.3 
& -0.93264 & 699  & 1 & 1 & 50  \\
54061.42434 & 132.3749 & 288.4474 & 30.7 & 580.4 & 161.71 & 551.1 & 184.5 
& -0.98450 & 515  & 1 & 1 & 100 \\
54103.32199 & 134.6975 & 294.9159 & 44.7 & 1066.4 & 163.89 & 1024.6 & 299.0 
& -0.98780 & 182  & 1 & 1 & 50  \\
\enddata
\tablecomments{
All quantities are in the ICRS 2000.0 reference frame.  The uncertainty 
values presented in these data have all been scaled by a factor of 1.73 
over the formal (internal) uncertainties within each given night.  
Column 1 is the heliocentric modified Julian date. Columns 2 and 3 
is the differential right ascension and declination between A and B, 
in milli-arcseconds.  Columns 4 and 5 are the $1\sigma$ uncertainties in 
the minor and major axis of the measurement uncertainty ellipse, in 
micro-arcseconds.  
Column 6, $\phi_{\rm e}$, is the angle between the major axis of the uncertainty 
ellipse and the right ascension axis, measured from increasing differential 
right ascension through increasing differential declination (the position 
angle of the uncertainty ellipse's orientation is $90-\phi_{\rm e}$).  
Columns 7 and 8 are the projected uncertainties in the right ascension and 
declination axis, in micro-arcseconds, while column 9 is the covariance 
between these.  Column 10 
is the number of scans taken during a given night.  Column 11 is 1 if the 
longitudinal dispersion compensator was in use, 0 otherwise.  Column 12 is 1 if 
the autoaligner was in use, 0 otherwise.  Column 13 represents the tracking 
frequency of the phase-referencing camera.  The quadrant was chosen such that 
the larger fringe contrast is designated the primary (contrast is a 
combination of source luminosity and interferometric visibility).
}
\end{deluxetable}

\subsection{Previous Measurements}

Previous differential astrometry measurements of $\mu$ Ori are tabulated in 
Table 5 of F2002.  These have been included in 
the current fit, with identical weightings as assigned by that investigation, 
though it is noted that the text contains a typographical error, and the 
$\rho$ unit uncertainty $\sigma_{\rho}$ should be 0.024, rather than 0.0024 
mas.  The time span of these measurements is much longer than that of the 
PHASES program and aids in solving the long period A-B orbit, which also lifts 
potential fit parameter degeneracies between that orbit and those of the 
short period subsystems.  Measurements marked as $3\sigma$ outliers by that 
investigation are omitted, resulting in 80 measurements each of separation and 
position angle being used for fitting.  Ten new measurements have been 
published since that investigation and are listed in Table 
\ref{muOriNewPrev} with weights computed with the same formula as used in 
F2002.  Two of these measurements are found to be $3\sigma$ outliers.  In 
total, there are 88 measurements of separation and position angle used in 
fitting.

\begin{deluxetable}{llllll}
\tablecolumns{6}
\tablewidth{0pc} 
\tablecaption{New Non-PHASES Astrometry for $\mu$ Ori\label{muOriNewPrev}}
\tablehead{
\colhead{Besselian Year} & \colhead{$\rho$} & \colhead{$\theta$} & \colhead{Weight} & \colhead{Outlier} & Reference
}
\startdata 
1991.8101 & 0.330 & 31.8  & 0.1  & 1 & \cite{TYC2002} \\
1996.8986 & 0.306 & 13.4  & 8.7  & 0 & \cite{Hor2001b} \\
1999.0153 & 0.200 & 14.8  & 10.1 & 1 & \cite{hor02} \\
1999.0153 & 0.196 & 13.3  & 9.7  & 1 & \cite{hor02} \\
1999.0153 & 0.203 & 14.9  & 10.4 & 1 & \cite{hor02} \\
1999.8915 & 0.150 & 11.7  & 5.8  & 1 & \cite{hor02} \\
1999.8915 & 0.145 & 12.5  & 5.5  & 1 & \cite{hor02} \\
1999.8915 & 0.154 & 11.3  & 6.1  & 1 & \cite{hor02} \\
2000.7653 & 0.081 & 359.4 & 2.0  & 0 & \cite{hor02} \\
2005.1331 & 0.179 & 26.4  & 0.2  & 1 & \cite{Sca2007a} \\
\enddata
\tablecomments{
The 10 new astrometry measurements published since F2002 for $\mu$ Ori.  
Column 1 is the epoch of observation in years, column 2 is the A-B separation 
in arcseconds, column 3 is the position angle east of north in degrees, 
and column 4 is the measurement weight on the same scale as F2002.  
($\delta$RA $=\rho \sin \theta$, $\delta$Dec $=\rho \cos \theta$.)  
Column 5 is 0 if the 
measurement is a $3\sigma$ outlier not used in fitting, 1 otherwise.  Column 
6 is the original work where the measurement was published.}
\end{deluxetable}

F2002 also present radial velocity observations of components Aa, 
Ba, and Bb.  Those measurements are included in the present fit, with 
weightings as reported in Tables 2, 3, and 4 of that paper.  Measurements 
marked as $3\sigma$ outliers by that investigation have not been included in 
the present analysis.  In total, 442 velocities---220 for Aa and 111 for each 
of Ba and Bb---are used in fitting.

\section{Orbital Models}

The apparent motions of the centers-of-light (photocenters) 
of the ${\rm A=Aa-Ab}$ and 
${\rm B=Ba-Bb}$ subsystems relative to each other are given by the model 
\begin{eqnarray}\label{muOri3DorbitEquation}
\overrightarrow{y_{\rm{obs}}} &=& 
\overrightarrow{r_{\rm{A-B}}} \nonumber\\
 &+& \frac{M_{\rm{Ab}}/M_{\rm{Aa}} - L_{\rm{Ab}}/L_{\rm{Aa}}}{\left(1+M_{\rm{Ab}}/M_{\rm{Aa}}\right)\left(1+L_{\rm{Ab}}/L_{\rm{Aa}}\right)}\overrightarrow{r_{\rm{Aa-Ab}}}\nonumber\\
 &-& \frac{M_{\rm{Bb}}/M_{\rm{Ba}} - L_{\rm{Bb}}/L_{\rm{Ba}}}{\left(1+M_{\rm{Bb}}/M_{\rm{Ba}}\right)\left(1+L_{\rm{Bb}}/L_{\rm{Ba}}\right)}\overrightarrow{r_{\rm{Ba-Bb}}}
\end{eqnarray}
corresponding to a four-body hierarchical dynamical system (HDS).  
The quantities $M$ are component masses and $L$ are component 
luminosities, and each of the summed vectors is determined by a 2-body 
Keplerian model.  This model is used to fit the astrometric data; note that 
the total masses from the Aa-Ab and Ba-Bb orbits also show up as component 
masses for the A-B orbit, linking them, and that the mass ratios and 
luminosity ratios appear as additional parameters, degenerate with each 
other.  The radial velocities are fit by a simple superposition of individual 
Keplerians; these determine mass ratios, and the luminosity ratios become 
nondegenerate parameters.  The luminosity ratios are primarily 
constrained by the K-band PHASES data; the other astrometric data are not 
precise enough to detect the subsystem motions.

\begin{figure}[]
\plottwo{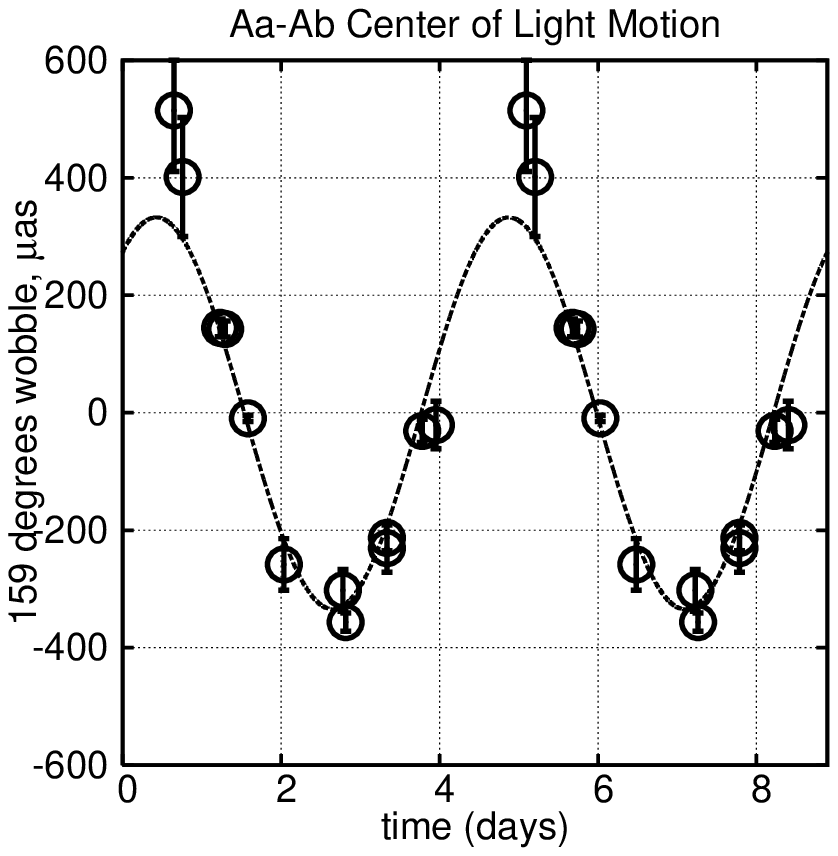}{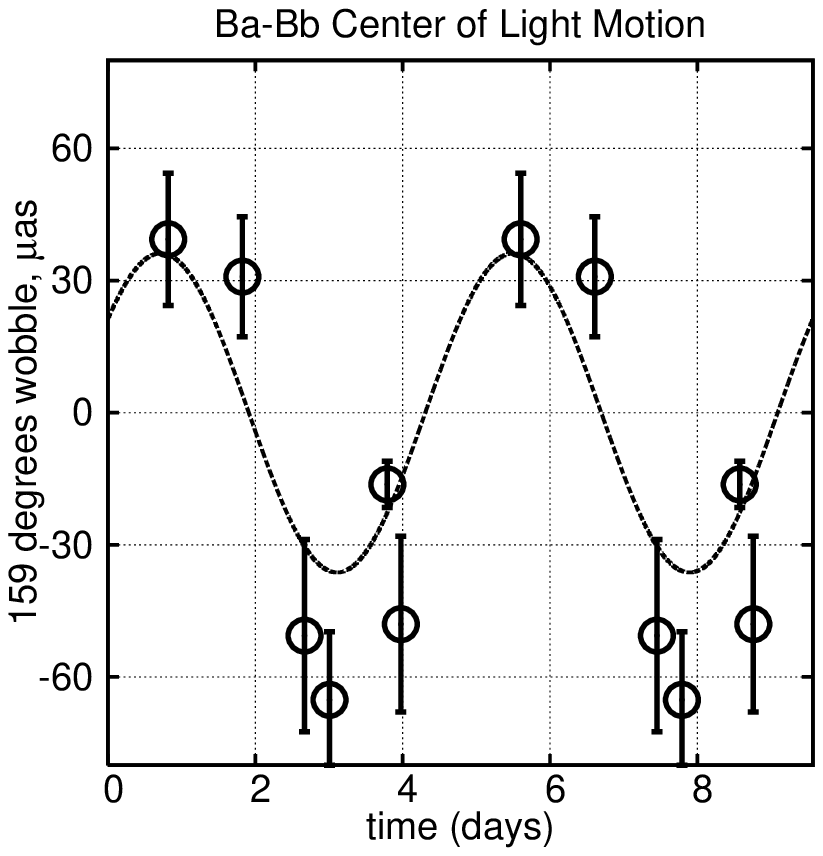}
\caption[The Orbit of $\mu$ Ori] 
{ \label{MuOriOrbit}
The astrometric orbits of $\mu$ Ori Aa-Ab and Ba-Bb, phase-wrapped about their 
respective orbital periods.  Phase zero is at the epoch of periastron passage 
($T$), and each plot is repeated for two cycles (and each measurement is 
plotted twice) to allow for continuity at all parts of the graph.  In both 
cases, the 
motions of the A-B system and of the other subsystem have been removed.
The projection axis shown for each is 159 degrees East of North (equivalent to 
position angle 291 degrees), well aligned with the minor axis of many PHASES 
observations.  The plotted uncertainties are those projected along this axis, 
and have been increased by a factor of 1.73 over the formal 
uncertainties, to reflect excess noise within the PHASES set.  On the left is 
the motion of Aa-Ab; 
for clarity, only those observations with rescaled and projected uncertainties 
less than 200 $\microas$ have been plotted; on the right is the motion of 
Ba-Bb, for which the cutoff was at 30 $\microas$.
}
\end{figure}

The combined simultaneous fit to all data sets has 26 free parameters and 
626 degrees of freedom.  The parameters used are listed in the top half of 
Table \ref{muOriOrbitModels} with their associated best-fit values and 
$1\sigma$ uncertainties.  It should be noted that while the Ba-Bb 
eccentricity has been allowed to vary as a free parameter, the best fit value 
is consistent with zero and could have been fixed; the other parameters are 
not changed significantly by doing so.  Quantities of interest derived from 
those parameters are listed in the second half of that table, with 
corresponding uncertainties derived from first-order uncertainty 
propagation.  The apparent center-of-light wobbles of the 
Aa-Ab and Ba-Bb subsystems are plotted in Figure \ref{MuOriOrbit}; the A-B 
and RV orbits were plotted in F2002 and are not significantly different in 
the present model.

Including PHASES measurements in the fit introduces the ability to 
evaluate the inclination and luminosity ratio of the Aa-Ab system and angles 
of the nodes of the Aa-Ab and Ba-Bb systems, quantities that were entirely 
unconstrained in the F2002 study.  The A-B angular parameters have much 
smaller uncertainties than in F2002 (the $\Omega_{\rm AB}$, $i_{\rm AB}$, and 
$\omega_{\rm AB}$ uncertainties are reduced by $13\times$, $6\times$, and 
$5\times$, respectively).  Uncertainties in the A-B period, eccentricity, and 
epoch of periastron passage are improved by a factor of 2 or more.  Most 
other fit parameters are constrained only marginally better than in F2002.

\begin{deluxetable}{lllllllll}
\tabletypesize{\tiny}
\tablecolumns{9}
\tablewidth{0pc} 
\tablecaption{Orbital parameters for $\mu$ Ori\label{muOriOrbitModels}}
\tablehead{
\colhead{} & \multicolumn{2}{c}{$L_{\rm Aa} > L_{\rm Ab}$,}
& \multicolumn{2}{c}{$L_{\rm Aa} > L_{\rm Ab}$,} 
& \multicolumn{2}{c}{$L_{\rm Aa} < L_{\rm Ab}$,} 
& \multicolumn{2}{c}{$L_{\rm Aa} < L_{\rm Ab}$,} \\
\colhead{Parameter} 
& \multicolumn{2}{c}{$L_{\rm Ba} > L_{\rm Bb}$}
& \multicolumn{2}{c}{$L_{\rm Ba} < L_{\rm Bb}$}
& \multicolumn{2}{c}{$L_{\rm Ba} > L_{\rm Bb}$}
& \multicolumn{2}{c}{$L_{\rm Ba} < L_{\rm Bb}$} \\
}
\startdata
$\chi^2$
& \multicolumn{2}{c}{723.86}
& \multicolumn{2}{c}{723.86}
& \multicolumn{2}{c}{723.58}
& \multicolumn{2}{c}{723.58} \\
$P_{\rm AB}$ (days)
& $6813.8 $ & $\pm 1.2$
& \nodata & \nodata    
& \nodata & \nodata    
& \nodata & \nodata \\ 
$e_{\rm AB}$                     
& $0.7410 $ & $\pm 0.0011$
& \nodata & \nodata    
& \nodata & \nodata    
& \nodata & \nodata \\ 
$i_{\rm AB}$ (degrees)
& $96.028 $ & $\pm 0.028$
& \nodata & \nodata    
& \nodata & \nodata    
& \nodata & \nodata \\ 
$\omega_{\rm AB}$ (degrees)
& $36.712 $ & $\pm 0.066$
& \nodata & \nodata    
& \nodata & \nodata    
& \nodata & \nodata \\ 
$T_{\rm AB}$ (MHJD)      
& $46090.7 $ & $\pm 1.0$
& \nodata & \nodata    
& \nodata & \nodata    
& \nodata & \nodata \\ 
$\Omega_{\rm AB}$ (degrees)
& $204.877 $ & $\pm 0.011$
& \nodata & \nodata    
& \nodata & \nodata    
& \nodata & \nodata \\ 
$P_{\rm AaAb}$ (days)
& $4.4475849 $ & $\pm 1.2\times 10^{-6}$
& \nodata & \nodata    
& \nodata & \nodata    
& \nodata & \nodata \\ 
$e_{\rm AaAb}$
& $0.0037 $ & $\pm 0.0014$
& \nodata & \nodata    
& \nodata & \nodata    
& \nodata & \nodata \\ 
$i_{\rm AaAb}$ (degrees)
& $47.1 $ & $\pm 9.0$
& \nodata & \nodata    
& $50.0 $ & $\pm 8.1$
& \nodata & \nodata \\ 
$\omega_{\rm AaAb}$ (degrees)
& $304 $ & $\pm 21$
& \nodata & \nodata    
& \nodata & \nodata    
& \nodata & \nodata \\ 
$T_{\rm AaAb}$ (MHJD)
& $43739.69 $ & $\pm 0.26$
& \nodata & \nodata    
& \nodata & \nodata    
& \nodata & \nodata \\ 
$\Omega_{\rm AaAb}$ (degrees)
& $50.5 $ & $\pm 3.7$
& \nodata & \nodata    
& $231.7 $ & $\pm 3.8$
& \nodata & \nodata \\ 
$P_{\rm BaBb}$ (days)
& $4.7835349 $ & $\pm 3.0\times 10^{-6}$
& \nodata & \nodata    
& \nodata & \nodata    
& \nodata & \nodata \\ 
$e_{\rm BaBb}$
& $0.0016 $ & $\pm 0.0014$
& \nodata & \nodata    
& \nodata & \nodata    
& \nodata & \nodata \\ 
$i_{\rm BaBb}$ (degrees)
& $110.71 $ & $\pm 0.73$
& \nodata & \nodata    
& \nodata & \nodata    
& \nodata & \nodata \\ 
$\omega_{\rm BaBb}$ (degrees)
& $217 $ & $\pm 47$
& \nodata & \nodata    
& \nodata & \nodata    
& \nodata & \nodata \\ 
$T_{\rm BaBb}$ (MHJD)
& $43746.40 $ & $\pm 0.63$
& \nodata & \nodata    
& \nodata & \nodata    
& \nodata & \nodata \\ 
$\Omega_{\rm BaBb}$ (degrees)
& $111.3 $ & $\pm 3.9$
& $291.3 $ & $\pm 3.9$
& $111.3 $ & $\pm 4.0$
& $291.3 $ & $\pm 4.0$ \\
$M_{\rm A}$ ($\Msun$)
& $3.030 $ & $\pm 0.069$
& \nodata & \nodata    
& \nodata & \nodata    
& \nodata & \nodata \\ 
$M_{\rm B}$ ($\Msun$)
& $2.746 $ & $\pm 0.038$
& \nodata & \nodata    
& \nodata & \nodata    
& \nodata & \nodata \\ 
$M_{\rm Ab}/M_{\rm Aa}$
& $0.274 $ & $\pm 0.051$
& \nodata & \nodata    
& $0.259 $ & $\pm 0.039$
& \nodata & \nodata \\ 
$M_{\rm Bb}/M_{\rm Ba}$
& $0.9764 $ & $\pm 0.0022$
& \nodata & \nodata    
& \nodata & \nodata    
& \nodata & \nodata \\ 
$L_{\rm Ab}/L_{\rm Aa}$
& $0 $ & $\pm 0.040$
& \nodata & \nodata    
& $0.738 $ & $\pm 0.061$
& \nodata & \nodata \\ 
$L_{\rm Bb}/L_{\rm Ba}$
& $0.765 $ & $\pm 0.055$
& $1.246 $ & $\pm 0.089$
& $0.773 $ & $\pm 0.055$
& $1.233 $ & $\pm 0.088$ \\
$d$ (parsecs)
& $46.11 $ & $\pm 0.28$
& \nodata & \nodata    
& \nodata & \nodata    
& \nodata & \nodata \\ 
$V_{0}$ (${\rm km\,s^{-1}}$)
& $42.548 $ & $\pm 0.027$
& \nodata & \nodata    
& \nodata & \nodata    
& \nodata & \nodata \\ 
\tableline
$M_{\rm Aa}$ ($\Msun$)
& $2.38 $ & $\pm 0.11 $
& \nodata & \nodata    
& $2.408 $ & $\pm 0.092 $
& \nodata & \nodata \\ 
$M_{\rm Ab}$ ($\Msun$)
& $0.652 $ & $\pm 0.097 $
& \nodata & \nodata    
& $0.623 $ & $\pm 0.075 $
& \nodata & \nodata \\ 
$M_{\rm Ba}$ ($\Msun$)
& $1.389 $ & $\pm 0.019 $
& \nodata & \nodata    
& \nodata & \nodata    
& \nodata & \nodata \\ 
$M_{\rm Bb}$ ($\Msun$)
& $1.356 $ & $\pm 0.019 $
& \nodata & \nodata    
& \nodata & \nodata    
& \nodata & \nodata \\ 
$\Phi_{\rm AB-AaAb}$ (degrees)
& $136.7 $ & $\pm 8.3 $
& \nodata & \nodata    
& $52.2  $ & $\pm 6.1 $
& \nodata & \nodata \\ 
$\Phi_{\rm AB-BaBb}$ (degrees)
& $91.2 $ & $\pm 3.6 $
& $84.5 $ & $\pm 3.6 $
& $91.2 $ & $\pm 3.8 $
& $84.5 $ & $\pm 3.8 $ \\
$\Phi_{\rm AaAb-BaBb}$ (degrees)
& $84.6 $ & $\pm 4.9 $
& $125.1 $ & $\pm 6.0 $ 
& $126.2 $ & $\pm 5.9 $
& $82.2 $ & $\pm 4.8 $ \\
$a_{\rm AB}$ (mas)
& $273.7 $ & $\pm 2.1$
& \nodata & \nodata    
& \nodata & \nodata    
& \nodata & \nodata \\ 
$a_{\rm AB}$ (AU)
& $12.620 $ & $\pm 0.057$
& \nodata & \nodata    
& \nodata & \nodata    
& \nodata & \nodata \\ 
$a_{\rm AaAb, COL}$ ($\microas$)
& $358 $ & $\pm 84$
& \nodata & \nodata    
& $364 $ & $\pm 53$
& \nodata & \nodata  \\
$a_{\rm AaAb}$ (mas)
& $1.661 $ & $\pm 0.016$
& \nodata & \nodata    
& \nodata & \nodata    
& \nodata & \nodata \\ 
$a_{\rm AaAb}$ (AU)
& $0.07659 $ & $\pm 0.00058$
& \nodata & \nodata    
& \nodata & \nodata    
& \nodata & \nodata \\ 
$a_{\rm BaBb, COL}$ ($\microas$)
& $102 $ & $\pm 30$
& \nodata & \nodata    
& $98 $ & $\pm 30$
& \nodata & \nodata \\ 
$a_{\rm BaBb}$ (mas)
& $1.688 $ & $\pm 0.013$
& \nodata & \nodata    
& \nodata & \nodata    
& \nodata & \nodata \\ 
$a_{\rm BaBb}$ (AU)
& $0.07780 $ & $\pm 0.00036$
& \nodata & \nodata    
& \nodata & \nodata    
& \nodata & \nodata \\ 
$\pi$ (mas)
& $21.69 $ & $\pm 0.13$
& \nodata & \nodata    
& \nodata & \nodata    
& \nodata & \nodata \\ 
$K_{\rm Aa}$ 
& $1.03 $ & $\pm 0.26$
& \nodata & \nodata    
& $1.64 $ & $\pm 0.26$
& \nodata & \nodata \\ 
$K_{\rm Ab}$ 
& \multicolumn{2}{c}{$ > 4.58$} 
& \multicolumn{2}{c}{\nodata} 
& $1.96 $ & $\pm 0.27$
& \nodata & \nodata \\ 
$K_{\rm Ba}$ 
& $1.72 $ & $\pm 0.26$
& $1.99 $ & $\pm 0.26$
& $1.73 $ & $\pm 0.26$
& $1.98 $ & $\pm 0.26$ \\
$K_{\rm Bb}$ 
& $2.02 $ & $\pm 0.26$
& $1.75 $ & $\pm 0.26$
& $2.01 $ & $\pm 0.26$
& $1.75 $ & $\pm 0.26$ \\
$L_{K,\,{\rm Aa}}$ 
& $8.3 $ & $\pm 2.0$
& \nodata & \nodata    
& $4.8 $ & $\pm 1.2$
& \nodata & \nodata \\ 
$L_{K,\,{\rm Ab}}$ 
& $0 $ & $\pm 0.33 $
& \nodata & \nodata    
& $3.52 $ & $\pm 0.86$
& \nodata & \nodata \\ 
$L_{K,\,{\rm Ba}}$ 
& $4.4 $ & $\pm 1.1$
& $3.45 $ & $\pm 0.84$
& $4.4 $ & $\pm 1.1$
& $3.47 $ & $\pm 0.84$ \\
$L_{K,\,{\rm Bb}}$ 
& $3.35 $ & $\pm 0.82$
& $4.3 $ & $\pm 1.0$
& $3.38 $ & $\pm 0.82$
& $4.3 $ & $\pm 1.0$ \\
\enddata
\tablecomments{
Orbital parameters for $\mu$ Ori.  In the second, third, and fourth solutions, 
ellipses indicate a parameter that changes by less than two units in the last 
reported digit from the previous model.  In the combined fits, all parameter 
uncertainties have been increased by a factor of $\sqrt{\chi_r^2} = 1.08$ 
(though the $\chi_r^2$ of the combined fit is artificial due to rescaling the 
uncertainties of the individual data sets, this reflects the degree to which 
the data sets agree with each other).
The first solution is strongly preferred as it produces masses and 
luminosities that are correlated; the second is also possible because the 
stars Ba and Bb are very similar.  The third and fourth solutions require an 
unlikely luminosity for component Ab, given its mass, and are not preferred.
$a_{\rm COL}$ is the semimajor axis of the motion of the center-of-light of a 
subsystem, at K-band.  $L_{\rm X}/L_{\rm Y}$ is the K-band luminosity ratio 
between components X and Y.  $K$ is the K-band absolute magnitude, $L_K$ is 
K-band luminosity, in solar units.  For the first two solutions, 
the best fit solution yields $K_{\rm Ab}$ is infinite; a lower limit is 
determined by setting $L_{\rm Ab}/L_{\rm Aa}$ to the upper limit of its 
$1\sigma$ uncertainty range.
}
\end{deluxetable}

\subsection{Relative Orbital Inclinations}

In systems with three or more stars, studying the system coplanarity is of 
interest for understanding the formation and evolution of multiples 
\citep{Sterzik2002}.  To determine this without ambiguity, one must have both 
visual and RV orbital solutions for pairs of interest.  Previously, this was 
available only in six triples \citep[for a listing, see][]{Mut06_v819her}.  
The reason mutual inclination measurements have been rare is due to the 
observational challenges these systems present---RV signals are largest for 
compact pairs of stars, whereas astrometry prefers wider pairs.  The ``wide'' 
pair must be studied with RV and thus have an orbital period (and 
corresponding separation) as short as the two-component binaries that are
already challenging to visual studies.  The ``narrow'' pair is even smaller.  
The mutual inclination between two orbits is given by:
\begin{equation}\label{KapPegMutualInclination}
\cos \Phi = \cos i_1 \cos i_2  + \sin i_1 \sin i_2 
\cos\left(\Omega_1 - \Omega_2\right)
\end{equation}
\noindent where $i_1$ and $i_2$ are the orbital inclinations and $\Omega_1$ 
and $\Omega_2$ are the longitudes of the ascending nodes.  If only 
velocities are 
available for one system, the orientation of that orbit is unknown (even if it 
is eclipsing, the longitude of the ascending node is unknown).  If velocities 
are unavailable for a given orbit, there is a degeneracy in which node is 
ascending---two values separated by 180 degrees are possible.  This gives two 
degenerate solutions for the mutual inclination (not necessarily separated by 
180 degrees).

Even when a center-of-light astrometric orbit is available for the narrow 
pair, there can be a degeneracy between which node is ascending and the 
luminosity ratio.  Having found one possible luminosity ratio 
$L_{\rm Ab}/L_{\rm Aa} = L_1$, it can be shown that the other possible 
solution, corresponding to changing the ascending node by 180 degrees, is 
given by
\begin{equation}\label{eqLum}
L_2 = \frac{2R+RL_1-L_1}{1+2L_1-R}
\end{equation}
\noindent where $R$ is the mass ratio $M_{\rm Ab}/M_{\rm Aa}$.  In previous 
studies, support data has been available to lift that degeneracy.  For 
example, in the V819 Her system \citep{Mut06_v819her}, the two possible 
luminosity ratios were $0.26$ and $1.89$.  The eclipsing nature of the Ba-Bb 
pair lifted that degeneracy because it helped establish that the luminosity 
ratio is much less than 1.  Similarly, in the $\kappa$ Peg system 
\citep{Mut06_kappeg}, the spectra used for RV also show that component Bb is 
much fainter than Ba, again lifting the degeneracy (a luminosity ratio either 
nearly zero or 1.9 were both possible).

For a quadruple system, there are as many as four degenerate fit solutions.  
For $\mu$ Ori, a global minimum $\chi^2$ is found with longitude of the 
ascending node $\Omega_{\rm AaAb} = 231.7 \pm 3.8$ degrees and 
$L_{\rm Ab}/L_{\rm Aa} = 0.738 \pm 0.061$.  The alternative pair of these 
parameters solving eq.~\ref{eqLum} would force the luminosity ratio to 
a negative value (but close to zero within uncertainties).  However, searching 
for solutions with nonnegative luminosity ratios near zero yields a fit 
solution with only slightly larger $\chi^2$, and the node at roughly 180 
degrees difference ($\Omega_{\rm AaAb} = 50.5 \pm 3.7$ and 
$L_{\rm Ab}/L_{\rm Aa} = 0 \pm 0.040$).  All parameters other than the node 
angle and luminosity ratio vary between the two models by amounts less than 
the fit uncertainties.  The two solutions find 
$M_{\rm Ab}/M_{\rm Aa} = 0.259 \pm 0.039$ or $ 0.274 \pm 0.051$ respectively; 
given the rough scaling $L \propto M^4$ \citep{MassLuminosity}, 
it is very likely that the larger luminosity ratio is 
incorrect.  The larger luminosity ratio would also imply Ab is as bright as 
Ba or Bb.  However, component Ab is not observed in the spectrum while Ba and 
Bb are.  This is suggestive that Ab is faint, though it is also possible to 
explain this lack of Ab lines by postulating it is rapidly rotating.  However, 
given that the other stars are not rapid rotators, there is little evidence to 
support Ab as a rapid rotator.  It is concluded that the luminosity ratio near 
zero is {\em strongly} preferred, despite the slightly worse $\chi^2$ fit.  
Both fits are reported in Table \ref{muOriOrbitModels}, but the rest of the 
discussion in this paper refers only to the preferred solution for Aa-Ab.  
This degeneracy can be fully lifted by a single epoch image with a closure 
phase capable interferometer with sufficient angular resolution 
\citep[such as the Navy Prototype Optical Interferometer,][]{arm98}.

For each of the Aa-Ab solutions, there exists two solutions for the Ba-Bb 
pair.  In these cases, no negative luminosity ratios are found; the degeneracy 
is perfect and the $\chi^2$ of the fits are identical.  $\Omega_{\rm BaBb}$ 
differs by 180 degrees in the two orbits, and the luminosity ratio 
$L_{\rm Bb}/L_{\rm Ba}$ switches between being larger (at 
$\Omega_{\rm BaBb} = 291$ degrees) or smaller (at 111 degrees) than unity; 
all other parameters remain unchanged.  Both solutions are near unity, and 
the stars have very similar masses 
($M_{\rm Bb}/M_{\rm Ba} = 0.9764 \pm 0.0022$).  While the solution for which 
Bb is less luminous than Ba is slightly more consistent because it correlates 
to the mass ratio, it is conceivable that the other solution is correct.  
Thus, the solution for which Bb is less luminous than Ba is slightly 
preferred, but not as conclusively as for the Aa-Ab case, where the 
differences between the stars are more significant.  Thus, both possibilities 
are considered in the remainder of this paper.

\subsection{Evidence for Kozai Cycles with Tidal Friction?}

Of particular interest is the potential for Kozai oscillations between orbital 
inclination and eccentricity in the narrow pairs \citep{Kozai1962}, which can 
affect the orbital evolution of the system.  These occur independent of 
distances or component masses, with the only requirement being that the mutual 
inclination is between 39.2 and $180-39.2=140.8$ degrees.  Other effects that 
cause precession can increase the value of this critical angle.

\cite{Fabrycky2007} predict a buildup of mutual inclinations near 40 and 140 
degrees by the combined effects of Kozai Cycles with Tidal Friction (KCTF) 
for triples whose short-period subsystems have periods between 3 and 
10 days.  
The mutual inclination of $\mu$ Ori AB-AaAb is near 140 degrees, leading one 
to consider if this trend is starting to be seen.  The systems with 
unambiguous mutual inclinations break down as follows:

\begin{itemize}
\item {\bf Five systems are outside of the 3-10 day inner period range.}  
These systems do not meet the criteria to be included in testing the 
buildup prediction:
\begin{enumerate}
\item V819 Her ($\Phi = 26.3\pm 1.5$ degrees, 2.23 d; see \S \ref{updates}), 
\item Algol \citep[$\Phi = 98.8\pm 4.9$ degrees, 2.9 d; ][]{les93, PanAlgol}, 
\item $\eta$ Vir \citep[$\Phi = 30.8\pm 1.3$ degrees, 72 d; ][]{hum03}, 
\item $\xi$ Uma ABC \citep[$\Phi = 132.1$ degrees, 670 d; ][]{Heintz1996}, 
and 
\item $\epsilon$ Hya ABC 
\citep[$\Phi = 39.4$ degrees, 5500 d; ][]{Heintz1996}.
\end{enumerate}
These fall outside the 3-10 day range of inner-binary periods applicable to 
the prediction in \cite{Fabrycky2007}.  However, it is worth noting the 
mutual inclinations of $\xi$ UMa and $\epsilon$ Hya {\em are} near 140 and 40 
degrees respectively and in V819 Her and $\eta$ Vir the values are outside 
the $40-140$ degrees range, so 
neither would have been predicted to undergo Kozai cycles or KCTF.  Algol has 
a nearly perpendicular alignment, though the dynamics of Algol are different 
due to quadrupole distortions in the semidetached  stars; Algol's alignment 
has been explained by \cite{eggleton2001}.
\item {\bf No systems are outside the 40-140 degrees range, while also in the 
3-10 day inner period range.}
\item {\bf Two systems are between 40 and 140 degrees and in the 3-10 day 
inner period range.}
These systems would not support the KCTF-driven buildup near 40 and 140 
degrees:
\begin{enumerate}
\item $\mu$ Ori AB-BaBb ($\Phi = 91.2 \pm 3.6$ or $84.5 \pm 3.6$ 
degrees are possible, 4.78 d) and 
\item 88 Tau AaAb-Ab1Ab2 
\citep[$\Phi = 82.0 \pm 3.3$ or $58 \pm 3.3$ degrees are possible, 7.89 d; ][]
{lane88Tau2007_draft}.
\end{enumerate}
While mutual inclination degeneracies continue to exist in both systems, all 
possible values are in this range.
\item {\bf Three systems are near 40 or 140 degrees, and in the 3-10 day 
inner period range.}
These systems would appear to support the KCTF prediction:
\begin{enumerate}
\item $\mu$ Ori AB-AaAb ($\Phi = 136.7 \pm 8.3$ degrees, 4.45 d), 
\item 88 Tau AaAb-Aa1Aa2 
\citep[$\Phi = 143.3 \pm 2.5$ degrees, 3.57 d; ][]{lane88Tau2007_draft}, and
\item $\kappa$ Peg 
($\Phi = 43.4\pm 3.9$ degrees, 5.97 d; see \S \ref{updates}).
\end{enumerate}
\end{itemize}

In total, 3 of the 5 systems meeting the criteria for testing the KCTF 
prediction do appear near the peak points of 40 and 140 degrees.  
Following equations 1, 22, and 35 in \cite{Fabrycky2007}, in the presence of 
general relativity (GR) precession one expects the critical angles for $\mu$ 
Ori AB-AaAb and $\kappa$ Peg to be increased from 39.2 degrees to $\sim 68$ 
($\tau \dot{\omega}_{GR} = 2.3$) and 
$\sim 54$ degrees ($\tau \dot{\omega}_{GR} = 1.3$), 
respectively, while for 88 Tau AaAb-Aa1Aa2 GR precession 
dominates no matter the inclination ($\tau \dot{\omega}_{GR} = 17$).  
Thus, Kozai oscillations are suppressed 
by precession in these systems' current states.  However, it is possible these 
were present at earlier stages in the systems' histories and their current 
configurations were still reached through KCTF---the inner binaries may have 
originally been more widely separated, in which case GR effects would have 
been reduced.

Alternatively, $\mu$ Ori AB-BaBb and 88 Tau AaAb-Ab1Ab2 both lie well within 
the range of predicted Kozai cycles, even including GR precession (which 
raises the critical angles to $\sim 58$ ($\tau \dot{\omega}_{GR} = 1.6$) and 
$\sim 50$ degrees ($\tau \dot{\omega}_{GR} = 0.9$), respectively).  

Several more systems with double visual orbits but lacking RV for at least 
one subsystem component are listed by \cite{Sterzik2002}.  Two more 
(HD 150680 and HD 214608) are mentioned by \cite{orlov2000} and one 
more (HD 108500) by \cite{orlov2005}.  Of these, 
the inner pair in HD 150680 is doubtful and listing HD 214608 as having a 
double visual orbit appears to be in error.  In the 
paper cited by \cite{orlov2000} for HD 214608, \cite{duq1987} reveal 
it to be a double spectroscopic system, but point out that the the visual 
elements of the inner pair are unconstrained.  The value of 150 degrees for 
the node seems to have been taken as nominal from the outer system (whose true 
ascending node is 180 degrees different).  Because the nodes cannot be 
distinguished in the visual-only pairs, two values of the mutual inclinations 
are equally possible for each.  Furthermore, all have inner systems with 
periods longer than 300 days.  Thus, these cannot provide further direction 
on testing the KCTF prediction.

\subsection{Masses and Distance}

Components Ba and Bb are each determined to $1.4\%$, Aa to $5\%$, while the 
lowest mass member, Ab, is uncertain at the $15\%$ level.  The individual 
masses of Aa and Ab have not been previously determined; this study enables 
the exploration of the natures of those stars.  Both are members of star 
classes of interest:  Aa is of spectral type Am, and Ab has a mass in the 
range of late K dwarfs.  Through their physical association, it can 
be assumed both are co-evolved with Ba and Bb, each within the mass range for 
which stellar models have been well calibrated through observation.

The masses of Ba and Bb are determined only slightly better (less than a 
factor of 2 improvement) over the previous study by F2002.  
Similarly, the distance is determined to $0.6\%$, a slight improvement.
The {\em Hipparcos} \citep{hipcat} based parallax values discussed 
by \cite{Soder1999} ($21.5\pm 0.8$ mas in the original evaluation, revised to 
$20.8 \pm 0.9$ mas when the binary nature was considered) are consistent with, 
but less well constrained, than the current value of $21.69 \pm 0.13$ mas.

\subsection{Component Luminosities}

The 2MASS K-band magnitude for $\mu$ Ori is $m_{\rm Total} = 3.637\pm 0.260$ 
\citep{2MASS}.  F2002 gives the difference between the luminosities 
of A and B in several bands.  Unfortunately, none of these were taken near 
the K band (2.2 $\mu$m) where PTI operates.  However, a Keck adaptive optics 
image of $\mu$ Ori was obtained on MJD 53227 with a narrow band 
${\rm H_2}$ 2-1 filter centered at 2.2622 microns.  The A-B differential 
magnitude in this band is $m_A - m_B = \Delta m_{AB} = -0.073 \pm 0.007$ 
magnitudes; this measurement is reported for the first time here.

The combined orbital fit provides the system distance $d = 46.11 \pm 0.28$ 
parsecs and the luminosity ratios.  The combined set of 
$m_{\rm Total}$, $\Delta m_{AB}$, $d$, $L_{\rm Ab}/L_{\rm Aa}$, and
$L_{\rm Bb}/L_{\rm Ba}$ determines the component luminosities.  Using first 
order error propagation, the K-band luminosities are 
$L_{K,\,{\rm Aa}} = 8.3 \pm 2.0$ solar and less than a third solar for Ab.  
Components Ba and Bb have K-band luminosities of either 
$L_{K,\,{\rm Ba}} = 4.4 \pm 1.1$ and $L_{K,\,{\rm Bb}} = 3.35 \pm 0.82$ or 
$L_{K,\,{\rm Ba}} = 3.45 \pm 0.84$ and $L_{K,\,{\rm Bb}} = 4.3 \pm 1.0$ solar.
Absolute magnitudes are also given in Table \ref{muOriOrbitModels}.  In each 
case, the uncertainty in the apparent magnitude $m_{\rm Total}$ dominates.

\subsection{System Age and Evolutionary Tracks}

\begin{figure}[]
\epsscale{0.5}
\plotone{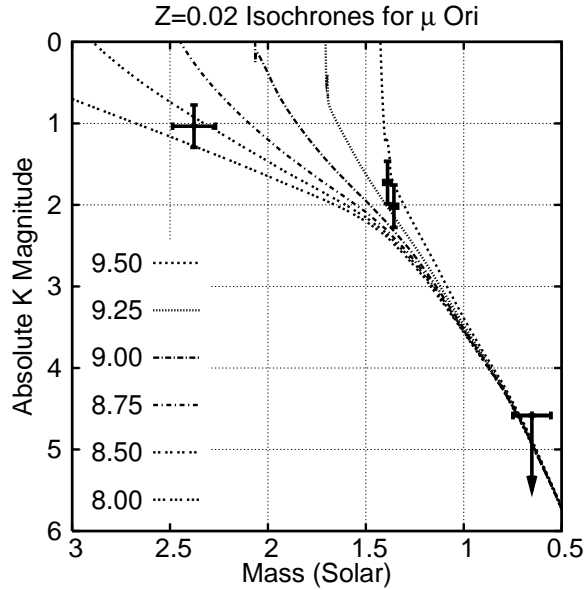}
\caption[$\mu$ Ori Isochrones] 
{ \label{MuOriIsochrones}
Stellar evolution models predicting isochrones for the elements of the $\mu$ 
Ori system show disagreement between component Aa with Ba and Bb.  Curve 
labels give values of ${\rm \log (Age/years)}$.  Component 
Aa provides the most leverage for determining the system age, though its 
evolution may have been altered by tidal friction.
}
\end{figure}

The masses and absolute K-band magnitudes for the components in this system 
can be compared to the stellar evolution models from \cite{Girardi2002}.  As 
in F2002, an abundance of $Z=0.02$ is assumed.  Figure \ref{MuOriIsochrones} 
shows the mass vs.~K-band magnitudes for several isochrones downloaded from 
http://pleiadi.pd.astro.it/ and the values derived for the components of 
$\mu$ Ori.  As the most massive component, the properties of Aa provide the 
strongest constraints on age, being most consistent with isochrones in the 
age range of $10^8-10^{8.5}$ years.  This is not entirely consistent with the 
properties of Ba and Bb, though close.  Of course, if KCTF has played a 
significant role in the orbital evolution of this system, one wonders whether 
stellar evolution models for single stars are really applicable to these 
stars.  One would anticipate the evolution of Aa as being affected by tidal 
forces because it is part of the subsystem near the predicted 140 degree 
``pile-up''.  Thus, one would rely more on Ba and Bb for system age 
determination, indicating an age over $10^9$ years.

\section{Updated Orbits}\label{updates}

During the course of this investigation, a sign error was found in the 
analysis software that was used to compute orbital solutions 
for $\delta$ Equulei \citep{Mut05_delequ}, $\kappa$ Pegasi 
\citep{Mut06_kappeg}, and V819 Herculis \citep{Mut06_v819her}.  This error 
affected 
fits to the radial velocity data only, with the result that the descending 
node was misidentified as the ascending, and the angle of periastron passage 
is off by 180 degrees.  Because the software was self-consistent, this has no 
impact on the mutual inclinations derived.  Additionally, the finite travel 
time of light across the wide orbit is included in the analysis, and this 
amount was thus incorrect by the same sign error, but the light travel time 
correction has only a small impact on those models.  Both errors have since 
been corrected.

Twenty-three new observations of $\delta$ Equ and 11 of $\kappa$ Peg have been 
collected since those initial investigations and are presented in Table 
\ref{phasesDataOthers}.  Also presented in Table \ref{phasesDataOthers} is the 
complete set of 34 V819 Her PHASES observations with the 10 measurements taken 
during eclipses marked; the previous investigation used less precise methods 
for predicting eclipse times, so the flagged measurements have changed.  
Measurements made during eclipse are not used in fitting.  

The new analysis 
makes use of the V819 Her Ba-Bb inclination constraint derived by eclipse 
lightcurves, a feature not included in the previous study.  When computing 
$\chi^2$, an additional term 
$(i_{\rm BaBb} - i_{\rm BaBb, \, eclipse})^2 / 
\sigma^2_{i {\rm , \, BaBb, \, eclipse}}$ is added to the sum, where 
$i_{\rm BaBb, \, eclipse} = 80.63$ and 
$\sigma_{i {\rm , \, BaBb, \, eclipse}} = 0.33$ degrees are the value and 
uncertainty of the Ba-Bb inclination from the lightcurve studies of 
\cite{vanHamme1994}.  Note that this measurement results from an entirely 
independent data set.  This added constraint lessens covariances between 
orbital elements.

The corrected and updated orbital solutions are presented in Tables 
\ref{delEquOrbitModels} and \ref{OthersOrbitModels}, which are fit to the 
complete set of PHASES observations and the other astrometric and RV 
observations tabulated in those previous papers.  (A few new measurement from 
speckle interferometry are available for each system since those 
investigations.  These have little impact on the orbital solutions are not 
included in the present fit to avoid over-complicating this update.)
The updated reweighting 
factors for the PHASES uncertainties for each data set are 3.91 for $\delta$ 
Equ, 7.93 for $\kappa$ Peg, and 2.0 for V819 Her.  Alternatively, the noise 
floor for $\kappa$ Peg is now found at 161 $\microas$.


\begin{deluxetable}{lllllllllllllll}
\tabletypesize{\tiny}
\tablecolumns{15}
\tablewidth{0pc} 
\tablecaption{New PHASES data for $\delta$ Equ, $\kappa$ Peg, and V819 Her\label{phasesDataOthers}}
\tablehead{ 
\colhead{Star} & \colhead{HJD-} & \colhead{$\delta$RA}    
& \colhead{$\delta$Dec}  & \colhead{$\sigma_{\rm min}$} 
& \colhead{$\sigma_{\rm maj}$} & \colhead{$\phi_{\rm e}$} 
&  \colhead{$\sigma_{\rm RA}$} & \colhead{$\sigma_{\rm Dec}$} 
& \colhead{$\frac{\sigma_{\rm RA, Dec}^2}{\sigma_{\rm RA}\sigma_{\rm Dec}}$} 
& \colhead{N} & \colhead{LDC} & \colhead{Align} & \colhead{Rate} & \colhead{Eclipse} \\
\colhead{}     & \colhead{2400000.5}             &  \colhead{(mas)}        
& \colhead{(mas)}        & \colhead{($\microas$)}       
& \colhead{($\microas$)}       & \colhead{(deg)}          
& \colhead{($\microas$)}       & \colhead{($\microas$)}     
& \colhead{} & \colhead{}  & \colhead{}    & \colhead{}      &  \colhead{(Hz)} &  \colhead{}
}
\startdata
$\delta$ Equ & 53508.50939 & -86.2658 & -117.6468 & 25.7 & 1872.8 
& 150.49 & 1630.0 & 922.7 & -0.99949 & 263 & 0 & 0 & 100 & 1 \\
$\delta$ Equ & 53550.41402 & -94.5630 & -143.7433 & 33.6 & 248.4 
& 151.98 & 219.8 & 120.4 & -0.94899 & 370   & 0 & 0 & 100 & 1 \\
$\delta$ Equ & 53552.38996 & -95.1309 & -144.7397 & 10.0 & 85.5 
& 150.13 & 74.3 & 43.5 & -0.96437 & 1352     & 0 & 0 & 100 & 1 \\
$\delta$ Equ & 53571.33634 & -99.1010 & -155.5409 & 9.8 & 68.1 
& 149.71 & 59.0 & 35.4 & -0.94762 & 1089      & 0 & 0 & 100 & 1 \\
$\delta$ Equ & 53584.32135 & -101.4805 & -162.5823 & 10.4 & 104.4 
& 151.49 & 91.9 & 50.7 & -0.97257 & 774    & 0 & 0 & 100 & 1 \\
$\delta$ Equ & 53586.30477 & -102.0695 & -163.5727 & 13.2 & 831.5 
& 150.94 & 726.8 & 404.0 & -0.99930 & 639  & 0 & 0 & 100 & 1 \\
$\delta$ Equ & 53607.23078 & -105.7105 & -174.9135 & 3.1 & 80.6 
& 148.50 & 68.8 & 42.2 & -0.99617 & 4856     & 0 & 0 & 100 & 1 \\
$\delta$ Equ & 53613.22582 & -107.1684 & -177.7125 & 6.7 & 243.8 
& 150.28 & 211.8 & 121.0 & -0.99794 & 1601  & 0 & 0 & 100 & 1 \\
$\delta$ Equ & 53614.22329 & -106.8636 & -178.5240 & 5.9 & 174.7 
& 150.30 & 151.8 & 86.7 & -0.99697 & 1597   & 0 & 0 & 100 & 1 \\
$\delta$ Equ & 53637.20185 & -111.2007 & -189.7814 & 12.3 & 357.4 
& 157.38 & 330.0 & 137.9 & -0.99531 & 570  & 0 & 0 & 100 & 1 \\
$\delta$ Equ & 53656.13045 & -114.1207 & -198.7179 & 16.3 & 215.1 
& 154.15 & 193.8 & 94.9 & -0.98163 & 526   & 0 & 0 & 100 & 1 \\
$\delta$ Equ & 53874.48665 & -133.8986 & -275.5257 & 12.3 & 424.2 
& 147.53 & 357.9 & 228.0 & -0.99794 & 264  & 0 & 0 & 100 & 1 \\
$\delta$ Equ & 53909.42854 & -134.0082 & -283.8970 & 6.1 & 58.2 
& 152.62 & 51.7 & 27.3 & -0.96785 & 2646     & 0 & 1 & 50 & 1 \\
$\delta$ Equ & 53957.31149 & -133.6026 & -293.0248 & 9.0 & 405.5 
& 154.61 & 366.4 & 174.1 & -0.99836 & 1127  & 1 & 1 & 50 & 1 \\
$\delta$ Equ & 53970.25612 & -133.8368 & -294.8599 & 5.7 & 50.6 
& 153.56 & 45.3 & 23.1 & -0.96133 & 2387     & 1 & 1 & 50 & 1 \\
$\delta$ Equ & 53971.25621 & -133.8811 & -294.9923 & 7.3 & 62.3 
& 152.11 & 55.1 & 29.8 & -0.96095 & 1682     & 1 & 1 & 50 & 1 \\
$\delta$ Equ & 53977.26493 & -133.7524 & -295.9196 & 5.3 & 39.3 
& 157.55 & 36.4 & 15.8 & -0.93151 & 3320     & 1 & 1 & 50 & 1 \\
$\delta$ Equ & 54003.21361 & -133.0033 & -299.3443 & 16.7 & 206.4 
& 161.12 & 195.4 & 68.6 & -0.96621 & 835   & 1 & 1 & 50 & 1 \\
$\delta$ Equ & 54005.19180 & -133.0818 & -299.4815 & 11.6 & 115.3 
& 157.21 & 106.4 & 45.9 & -0.96173 & 1614  & 1 & 1 & 50 & 1 \\
$\delta$ Equ & 54028.10922 & -132.3883 & -301.8627 & 13.3 & 406.9 
& 153.38 & 363.9 & 182.7 & -0.99668 & 632  & 1 & 1 & 50 & 1 \\
$\delta$ Equ & 54037.11296 & -132.9388 & -302.3034 & 11.0 & 319.6 
& 159.14 & 298.7 & 114.3 & -0.99468 & 556  & 1 & 1 & 100 & 1 \\
$\delta$ Equ & 54230.50355 & -116.0819 & -301.5397 & 13.1 & 508.2 
& 146.70 & 424.8 & 279.2 & -0.99843 & 933  & 1 & 1 & 50 & 1 \\
$\delta$ Equ & 54266.47236 & -111.8997 & -297.2144 & 13.2 & 112.0 
& 158.12 & 104.1 & 43.5 & -0.94550 & 1315  & 1 & 1 & 50 & 1 \\
$\kappa$ Peg & 53494.50786 & 104.6067 & 43.3687 & 13.6 & 645.9 
& 143.23 & 517.5 & 386.8 & -0.99904 & 702     & 0 & 0 & 100 & 1 \\
$\kappa$ Peg & 53586.36471 & 83.9658 & 55.9066 & 2.2 & 11.5 
& 166.51 & 11.2 & 3.4 & -0.75093 & 5532          & 0 & 0 & 100 & 1 \\
$\kappa$ Peg & 53637.29586 & 71.2323 & 62.9289 & 6.3 & 53.5 
& 173.82 & 53.2 & 8.5 & -0.66476 & 1639          & 0 & 0 & 100 & 1 \\
$\kappa$ Peg & 53921.45172 & 25.0960 & 85.9815 & 43.1 & 10007.6 
& 160.71 & 9445.7 & 3306.4 & -0.99990 & 615  & 0 & 1 & 50 & 1 \\
$\kappa$ Peg & 53963.35858 & -9.1846 & 98.3661 & 6.2 & 79.3 
& 165.54 & 76.8 & 20.7 & -0.95159 & 2244         & 1 & 1 & 100 & 1 \\
$\kappa$ Peg & 53978.32708 & -14.0804 & 99.4673 & 2.7 & 17.5 
& 162.83 & 16.8 & 5.8 & -0.87056 & 5863         & 1 & 1 & 50 & 1 \\
$\kappa$ Peg & 53995.24384 & -17.0621 & 101.1081 & 11.5 & 514.2 
& 159.38 & 481.3 & 181.5 & -0.99770 & 389    & 1 & 1 & 100 & 1 \\
$\kappa$ Peg & 54003.32618 & -22.4829 & 101.1592 & 10.9 & 849.5 
& 3.09 & 848.2 & 47.1 & 0.97271 & 1590       & 1 & 1 & 50 & 1 \\
$\kappa$ Peg & 54008.31093 & -21.6029 & 102.0991 & 4.6 & 114.1 
& 2.48 & 113.9 & 6.7 & 0.73308 & 2850         & 1 & 1 & 50 & 1 \\
$\kappa$ Peg & 54075.12667 & -38.5846 & 107.2133 & 7.0 & 183.8 
& 3.06 & 183.5 & 12.0 & 0.81528 & 1756        & 1 & 1 & 50 & 1 \\
$\kappa$ Peg & 54265.39966 & -84.8163 & 119.9685 & 8.2 & 389.0 
& 142.94 & 310.4 & 234.5 & -0.99904 & 3661    & 1 & 1 & 50 & 1 \\
V819 Her & 53109.47951 & 49.6406 & -84.4966 & 7.3 & 282.5 
& 158.77 & 263.4 & 102.5 & -0.99707 & 2011       & 0 & 0 & 100 & 1  \\
V819 Her & 53110.48183 & 48.0946 & -84.1334 & 11.9 & 600.4 
& 159.53 & 562.5 & 210.3 & -0.99819 & 1334       & 0 & 0 & 100 & 1  \\
V819 Her & 53123.45772 & 49.1860 & -85.9318 & 18.1 & 507.8 
& 162.47 & 484.3 & 153.9 & -0.99240 & 1378       & 0 & 0 & 100 & 0  \\
V819 Her & 53130.44208 & 48.4778 & -86.4135 & 6.6 & 205.8 
& 162.94 & 196.7 & 60.7 & -0.99360 & 2537        & 0 & 0 & 100 & 1  \\
V819 Her & 53137.43044 & 48.3928 & -87.1396 & 14.0 & 280.2 
& 164.34 & 269.9 & 76.8 & -0.98202 & 1226        & 0 & 0 & 100 & 1  \\
V819 Her & 53144.42426 & 47.7017 & -87.6612 & 25.1 & 1039.6 
& 167.13 & 1013.5 & 232.9 & -0.99386 & 897       & 0 & 0 & 100 & 1  \\
V819 Her & 53145.39541 & 48.3082 & -87.8013 & 13.7 & 316.5 
& 161.59 & 300.3 & 100.8 & -0.98964 & 1673       & 0 & 0 & 100 & 1  \\
V819 Her & 53168.33949 & 47.0275 & -89.7513 & 15.0 & 339.9 
& 162.93 & 325.0 & 100.8 & -0.98778 & 1409       & 0 & 0 & 100 & 1  \\
V819 Her & 53172.35221 & 47.3441 & -90.1337 & 6.3 & 170.0 
& 168.29 & 166.5 & 35.1 & -0.98309 & 2560        & 0 & 0 & 100 & 1  \\
V819 Her & 53173.33202 & 47.1604 & -90.3599 & 8.0 & 77.4 
& 33.97 & 64.4 & 43.8 & 0.97548 & 2904           & 0 & 0 & 100 & 1  \\
V819 Her & 53181.33391 & 46.4333 & -90.7857 & 7.5 & 174.6 
& 169.71 & 171.8 & 32.1 & -0.97114 & 2795        & 0 & 0 & 100 & 0  \\
V819 Her & 53182.33164 & 46.5646 & -91.0136 & 13.9 & 333.2 
& 169.62 & 327.8 & 61.6 & -0.97328 & 2014        & 0 & 0 & 100 & 1  \\
V819 Her & 53186.30448 & 45.6584 & -91.1213 & 18.2 & 426.3 
& 166.80 & 415.1 & 99.0 & -0.98197 & 706         & 0 & 0 & 100 & 1  \\
V819 Her & 53187.30462 & 46.1427 & -91.2225 & 13.0 & 441.5 
& 166.94 & 430.1 & 100.6 & -0.99110 & 1578       & 0 & 0 & 100 & 1  \\
V819 Her & 53197.26851 & 46.1852 & -92.1522 & 4.8 & 117.1 
& 164.87 & 113.0 & 30.9 & -0.98715 & 5218        & 0 & 0 & 100 & 0  \\
V819 Her & 53198.24258 & 46.2937 & -92.2555 & 5.7 & 54.6 
& 160.37 & 51.5 & 19.1 & -0.94836 & 5404         & 0 & 0 & 100 & 0  \\
V819 Her & 53199.29186 & 44.0252 & -91.9446 & 24.7 & 1645.4 
& 171.42 & 1627.0 & 246.7 & -0.99488 & 946       & 0 & 0 & 100 & 0  \\
V819 Her & 53208.25236 & 46.4303 & -92.4901 & 6.6 & 181.6 
& 37.67 & 143.8 & 111.1 & 0.99718 & 6558         & 0 & 0 & 100 & 0  \\
V819 Her & 53214.24077 & 45.6337 & -93.3429 & 5.5 & 125.9 
& 169.45 & 123.8 & 23.7 & -0.97194 & 5251        & 0 & 0 & 100 & 1  \\
V819 Her & 53215.23094 & 45.6364 & -93.5172 & 4.8 & 110.6 
& 167.53 & 108.0 & 24.3 & -0.97962 & 5723        & 0 & 0 & 100 & 1  \\
V819 Her & 53221.22209 & 46.2559 & -92.9726 & 8.8 & 342.2 
& 38.91 & 266.3 & 215.0 & 0.99860 & 3998         & 0 & 0 & 100 & 1  \\
V819 Her & 53228.20946 & 45.0884 & -94.3310 & 7.2 & 100.6 
& 169.45 & 98.9 & 19.7 & -0.92813 & 3180         & 0 & 0 & 100 & 0  \\
V819 Her & 53229.22073 & 45.2196 & -94.5160 & 6.4 & 80.0 
& 172.84 & 79.4 & 11.8 & -0.83914 & 3905         & 0 & 0 & 100 & 1  \\
V819 Her & 53233.18295 & 45.0993 & -94.8458 & 6.0 & 64.7 
& 167.67 & 63.2 & 15.0 & -0.91188 & 3303         & 0 & 0 & 100 & 1  \\
V819 Her & 53234.20151 & 44.8269 & -94.7637 & 7.6 & 37.8 
& 172.74 & 37.5 & 8.9 & -0.51352 & 3701          & 0 & 0 & 100 & 1  \\
V819 Her & 53235.21764 & 45.2153 & -94.9183 & 8.5 & 107.1 
& 176.57 & 106.9 & 10.6 & -0.60015 & 2094        & 0 & 0 & 100 & 1  \\
V819 Her & 53236.16733 & 44.4598 & -94.8862 & 4.7 & 78.1 
& 166.59 & 76.0 & 18.7 & -0.96552 & 6684         & 0 & 0 & 100 & 0  \\
V819 Her & 53249.16006 & 44.2467 & -95.8032 & 4.3 & 87.9 
& 172.71 & 87.2 & 12.0 & -0.93121 & 5428         & 0 & 0 & 100 & 1  \\
V819 Her & 53466.52265 & 31.4132 & -102.8881 & 9.9 & 204.3 
& 163.01 & 195.4 & 60.4 & -0.98524 & 3031        & 0 & 0 & 100 & 1  \\ 
V819 Her & 53481.50628 & 30.4432 & -103.3348 & 11.1 & 311.1 
& 38.18 & 244.6 & 192.5 & 0.99731 & 3301         & 0 & 0 & 100 & 0  \\ 
V819 Her & 53494.45305 & 29.5257 & -103.1671 & 18.6 & 251.9 
& 163.98 & 242.2 & 71.8 & -0.96316 & 1355        & 0 & 0 & 100 & 1  \\ 
V819 Her & 53585.24930 & 23.9818 & -102.6116 & 10.1 & 114.7 
& 174.02 & 114.0 & 15.6 & -0.76046 & 1479        & 0 & 0 & 100 & 1  \\ 
V819 Her & 53874.42827 & 1.2427 & -88.7776 & 6.8 & 99.5 
& 168.07 & 97.4 & 21.6 & -0.94665 & 2358         & 0 & 0 & 100 & 1  \\ 
V819 Her & 53956.21398 & -5.0810 & -81.9129 & 9.2 & 341.7 
& 170.23 & 336.8 & 58.7 & -0.98729 & 4028        & 1 & 1 & 50  & 0  \\ 
\enddata
\tablecomments{
All quantities are in the ICRS 2000.0 reference frame.  The uncertainty values 
presented in these data have not been rescaled.  
Column 1 is the star name.  Columns 2-14 are as columns 1-13 in Table 
\ref{phasesDataMuOri}.  Column 15 is 0 if the measurement was 
taken during a subsystem eclipse, 1 otherwise (V819 Her only).
}
\end{deluxetable}

\begin{deluxetable}{lll}
\tablecolumns{3}
\tablewidth{0pc} 
\tablecaption{Orbital parameters for $\delta$ Equ\label{delEquOrbitModels}}
\tablehead{
\colhead{Parameter} & \colhead{Value} & \colhead{Uncertainty}
}
\startdata 
$P$ (days)                         & 2084.03   & $\pm 0.10$     \\
$T$ (MHJD)                         & 53112.071 & $\pm 0.052$    \\
$e$                                & 0.436851  & $\pm 0.000025$ \\
$a$ (mas)                          & 231.9650  & $\pm 0.0080$   \\
$V_{0, Lick}$ (${\rm km\,s^{-1}}$) & $-15.40$  & $\pm 0.11$     \\
$V_{0, DAO}$ (${\rm km\,s^{-1}}$)  & $-15.875$ & $\pm 0.080$    \\
$V_{0, C}$ (${\rm km\,s^{-1}}$)    & $-15.73$  & $\pm 0.10$     \\
$M_1$ ($\Msun$)                    & 1.192     & $\pm 0.012$    \\
$M_2$ ($\Msun$)                    & 1.187     & $\pm 0.012$    \\
$M_1+M_2$ ($\Msun$)                & 2.380     & $\pm 0.019$    \\
$M_1/M_2$                          & 1.004     & $\pm 0.012$    \\
$i$ (deg)                          & 99.4083   & $\pm 0.0098$   \\
$\omega$ (deg)                     & 7.735     & $\pm 0.013$  \\
$\Omega$ (deg)                     & 23.362    & $\pm 0.012$    \\
$d$ (pc)                           & 18.379    & $\pm 0.048$    \\
\tableline
$\pi$ (mas)                        & 54.41     & $\pm 0.14$     \\
\enddata
\tablecomments{
All parameter uncertainties have been increased by a factor of 
$\sqrt{\chi_r^2} = 1.09$ (though the $\chi_r^2$ of the combined fit is 
artificial due to rescaling the uncertainties of the individual data sets, 
this reflects the degree to which the data sets agree with each other).
The fit was repeated several times varying the set of non-degenerate 
parameters used in order to obtain uncertainty estimates for a number of 
desired quantities.  The parameters $\{a, R=M_1/M_2\}$ were 
replaced with the sets $\{M=M_1+M_2, R\}$ and $\{M_1, M_2\}$.  
The parallax is a derived quantity.}
\end{deluxetable}

\begin{deluxetable}{lllll}
\tabletypesize{\footnotesize}
\tablecolumns{5}
\tablewidth{0pc} 
\tablecaption{Orbital parameters for $\kappa$ Peg and V819 Her\label{OthersOrbitModels}}
\tablehead{
\colhead{} & \multicolumn{2}{c}{$\kappa$ Peg} & \multicolumn{2}{c}{V819 Her} \\
\colhead{Parameter} & \colhead{Value} & \colhead{Uncertainty} & \colhead{Value} & \colhead{Uncertainty} 
}
\startdata 
$P_{AB}$ (days)                             
& 4224.76   & $\pm 0.74$      & 2019.66 & $\pm 0.35$  \\
$T_{AB}$ (MHJD)                             
& 52401.52  & $\pm 0.96$      & 52627.5 & $\pm 1.3$   \\
$e_{AB}$                                    
& 0.3140    & $\pm 0.0011$    & 0.67974 & $\pm 0.00066$ \\
$i_{AB}$ (degrees)                          
& 107.911   & $\pm 0.029$     & 56.40   & $\pm 0.13$  \\
$\omega_{AB}$ (degrees)                     
& 124.666   & $\pm 0.064$     & 222.50  & $\pm 0.22$  \\
$\Omega_{AB}$ (degrees)                     
& 289.037   & $\pm 0.021$     & 141.96  & $\pm 0.12$  \\
$P_{BaBb}$ (days)                           
& 5.9714971 & $\pm 1.3 \times10^{-6}$   
& 2.2296330 & $\pm 1.9 \times10^{-6}$ \\
$T_{BaBb}$ (MHJD)                           
& 52402.22  & $\pm 0.10$      & 52627.17 & $\pm 0.29$ \\
$e_{BaBb}$                                  
& 0.0073    & $\pm 0.0013$    & 0.0041   & $\pm 0.0033$ \\
$i_{BaBb}$ (degrees)                        
& 125.7     & $\pm 5.1$       & 80.70     & $\pm 0.38$  \\
$\omega_{BaBb}$ (degrees)                   
& 179.0     & $\pm 6.0$       & 227      & $\pm 47$   \\
$\Omega_{BaBb}$ (degrees)                   
& 244.1     & $\pm 2.3$       & 131.1    & $\pm 4.1$  \\
$V_{0, Keck}$ (${\rm km\,s^{-1}}$)           
& $-9.46$   & $\pm 0.22$      & \nodata  & \nodata    \\
$V_{1, Keck}$ (${\rm km\,s^{-1}\,day^{-1}}$) 
& $-2.2$ $\times 10^{-4}$ & $\pm 3.4 \times 10^{-4}$ & \nodata & \nodata \\
$V_{2, Keck}$ (${\rm km\,s^{-1}\,day^{-2}}$) 
& $6.8 \times 10^{-6}$  & $\pm 2.4 \times 10^{-6}$ & \nodata & \nodata \\
$V_{0, C}$ (${\rm km~s^{-1}}$)               
& $-9.40$   & $\pm 0.26$      & \nodata & \nodata     \\
$V_{0, Lick}$ (${\rm km~s^{-1}}$)            
& $-8.37$   & $\pm 0.26$      & \nodata & \nodata     \\
$V_{0, M/K}$ (${\rm km~s^{-1}}$)             
& \nodata   & \nodata         & $-3.375$ & $\pm 0.059$ \\
$V_{0, DAO}$ (${\rm km~s^{-1}}$)             
& \nodata   & \nodata         & $-3.385$ & $\pm 0.065$ \\
$V_{0, DDO}$ (${\rm km~s^{-1}}$)             
& \nodata   & \nodata         & $-3.35$ & $\pm 0.12$  \\
$M_A$ ($\Msun$)                              
& 1.533     & $\pm 0.050$     & 1.799   & $\pm 0.098$ \\
$M_{Ba+Bb}$ ($\Msun$)                        
& 2.472     & $\pm 0.078$     & 2.560   & $\pm 0.067$ \\
$M_{Bb}/M_{Ba}$                              
& 0.501     & $\pm 0.049$     & 0.742   & $\pm 0.012$ \\
$L_{Bb}/L_{Ba}$                              
& 0.015     & $\pm 0.021$     & 0.280   & $\pm 0.037$ \\
$d$ (parsecs)                                
& 34.57     & $\pm 0.21$      & 68.65   & $\pm 0.87$  \\
\tableline
$\Phi_{\rm AB-BaBb}$ (degrees)               
& 43.4      & $\pm 3.9$       & 26.3    & $\pm 1.5$   \\
$M_{Ba}$ ($\Msun$)                           
& 1.646     & $\pm 0.074$     & 1.469   & $\pm 0.040$ \\
$M_{Bb}$ ($\Msun$)                           
& 0.825     & $\pm 0.059$     & 1.090   & $\pm 0.030$ \\
$a_{AB}$ (AU)                               
& 8.122     & $\pm 0.063$     & 5.108   & $\pm 0.046$ \\
$a_{BaBb}$ (AU)                             
& 0.08710   & $\pm 0.00091$   & 0.04569 & $\pm 0.00040$ \\
$\pi$ (mas)                                  
& 28.93     & $\pm 0.18$      & 14.57   & $\pm 0.19$ \\
\enddata
\tablecomments{
Uncertainties for $\kappa$ Pegasi are the maximum of three uncertainties:  
the uncertainty from the combined fit that included PHASES-reweighted data, 
that including PHASES data with a $161\microas$ noise floor, and the 
difference in the fit values for the two models.  The parameters are the 
average values between a fit including the reweighted uncertainties and 
one with the noise floor.
}
\end{deluxetable}

\section{Conclusions}

The center-of-light astrometric motions of the Aa-Ab and Ba-Bb subsystems in 
$\mu$ Ori have been constrained by PHASES observations.  While four degenerate 
orbital solutions exist, two of these can be excluded with high reliability 
based on mass-luminosity arguments, and the fact that Ab is not observed in 
the spectra.  Ba and Bb are stars of a class (mid-F dwarfs) whose properties 
have been well established by studying other binaries.  Their association with 
Aa and Ab, which are members of more poorly studied classes (Am and late K 
dwarfs) allows a better understanding of those objects in a system which can 
be assumed to be coevolved.  The orbital solution finds masses and 
luminosities for all four components, the basic properties necessary in 
studying their natures.

Complex dynamics must occur in $\mu$ Ori.  The Ba-Bb orbital plane is nearly 
perpendicular to that of the A-B motion, and certainly undergoes Kozai-type 
inclination-eccentricity oscillations.  It is possible that the mutual 
inclination of the A-B pair and Aa-Ab subsystem is a result of KCTF effects 
over the system's evolution.

Finally, it is noted that the orbits in the $\mu$ Ori system are quite 
non-coplanar.  This is in striking contrast with the planets of the solar 
system, but follows the trend seen in triple star systems.  With the solar 
system being the only one whose coplanarity has been evaluated, it is 
difficult to draw conclusions about the configurations of planetary systems in 
general.  It is important that future investigations evaluate the 
coplanarities 
of extrasolar planetary systems to establish a distribution.  Whether that 
distribution will be the same or different than that of their stellar 
counterparts may point to similarities or differences in star and planet 
formation, and provide a key constraint on modeling multiple star and planet 
formation.

\acknowledgements 
We thank Daniel Fabrycky for helpful correspondence about his recent 
theoretical work on KCTF.  We thank Bill Hartkopf for providing weights for 
the new non-PHASES differential astrometry measurements.  
PHASES benefits from the efforts of the PTI collaboration members who have 
each contributed to the development of an extremely reliable observational 
instrument.  Without this outstanding engineering effort to produce a solid 
foundation, advanced phase-referencing techniques would not have been 
possible.  We thank PTI's night assistant Kevin Rykoski for his efforts to 
maintain PTI in excellent condition and operating PTI in phase-referencing 
mode every week.  Part of the work described in this paper was performed at 
the Jet Propulsion Laboratory under contract with the National Aeronautics 
and Space Administration. Interferometer data was obtained at the Palomar
Observatory with the NASA Palomar Testbed Interferometer, supported
by NASA contracts to the Jet Propulsion Laboratory.  This publication makes 
use of data products from the Two Micron All Sky Survey, which is a joint 
project of the University of Massachusetts and the Infrared Processing and 
Analysis Center/California Institute of Technology, funded by the National 
Aeronautics and Space Administration and the National Science Foundation.  
This research has made use of the Simbad database, operated at CDS, 
Strasbourg, France.  MWM acknowledges support from the Townes Fellowship 
Program.  PHASES is funded in part by the California Institute of Technology 
Astronomy Department, and by the National Aeronautics and Space Administration 
under Grant No.~NNG05GJ58G issued through the Terrestrial Planet Finder 
Foundation Science Program.  This work was supported in part by the National 
Science Foundation through grants AST 0300096, AST 0507590, and AST 005366.
The work of FCF has been supported in part by NASA grant NCC5-511 and 
NSF grant HRD-9706268.  MK is supported by the Polish Ministry of Science and 
Higher Education through grants N203 005 32/0449 and 1P03D 021 29.

\bibliography{main}

\newcommand{\noopsort}[1]{} \newcommand{\printfirst}[2]{#1}
  \newcommand{\singleletter}[1]{#1} \newcommand{\switchargs}[2]{#2#1}
\begin{thebibliography}{34}
\expandafter\ifx\csname natexlab\endcsname\relax\def\natexlab#1{#1}\fi
\expandafter\ifx\csname url\endcsname\relax
  \def\url#1{{\tt #1}}\fi

\bibitem[{Aitken}(1914)]{Aitken1914}
R.~G. {Aitken}.
\newblock {One hundred new double stars : twenty-second list}.
\newblock {\em Lick Observatory Bulletin}, 8:\penalty0 93--95, 1914.

\bibitem[{Armstrong} et~al.(1998){Armstrong}, {Mozurkewich}, {Rickard},
  {Hutter}, {Benson}, {Bowers}, {Elias}, {Hummel}, {Johnston}, {Buscher},
  {Clark}, {Ha}, {Ling}, {White}, and {Simon}]{arm98}
J.~T. {Armstrong}, D.~{Mozurkewich}, L.~J. {Rickard}, D.~J. {Hutter}, J.~A.
  {Benson}, P.~F. {Bowers}, N.~M. {Elias}, C.~A. {Hummel}, K.~J. {Johnston},
  D.~F. {Buscher}, J.~H. {Clark}, L.~{Ha}, L.-C. {Ling}, N.~M. {White}, and
  R.~S. {Simon}.
\newblock {The Navy Prototype Optical Interferometer}.
\newblock {\em \apj}, 496:\penalty0 550--571, March 1998.

\bibitem[{Colavita} et~al.(1999){Colavita}, {Wallace}, {Hines}, {Gursel},
  {Malbet}, {Palmer}, {Pan}, {Shao}, {Yu}, {Boden}, {Dumont}, {Gubler},
  {Koresko}, {Kulkarni}, {Lane}, {Mobley}, and {van Belle}]{col99}
M.~M. {Colavita}, J.~K. {Wallace}, B.~E. {Hines}, Y.~{Gursel}, F.~{Malbet},
  D.~L. {Palmer}, X.~P. {Pan}, M.~{Shao}, J.~W. {Yu}, A.~F. {Boden}, P.~J.
  {Dumont}, J.~{Gubler}, C.~D. {Koresko}, S.~R. {Kulkarni}, B.~F. {Lane}, D.~W.
  {Mobley}, and G.~T. {van Belle}.
\newblock {The Palomar Testbed Interferometer}.
\newblock {\em \apj}, 510:\penalty0 505--521, January 1999.

\bibitem[{Duquennoy}(1987)]{duq1987}
A.~{Duquennoy}.
\newblock {A study of multiple stellar systems with CORAVEL.}
\newblock {\em \aap}, 178:\penalty0 114--130, May 1987.

\bibitem[{Eggleton} and {Kiseleva-Eggleton}(2001)]{eggleton2001}
P.~P. {Eggleton} and L.~{Kiseleva-Eggleton}.
\newblock {Orbital Evolution in Binary and Triple Stars, with an Application to
  SS Lacertae}.
\newblock {\em \apj}, 562:\penalty0 1012--1030, December 2001.

\bibitem[{Fabricius} et~al.(2002){Fabricius}, {H{\o}g}, {Makarov}, {Mason},
  {Wycoff}, and {Urban}]{TYC2002}
C.~{Fabricius}, E.~{H{\o}g}, V.~V. {Makarov}, B.~D. {Mason}, G.~L. {Wycoff},
  and S.~E. {Urban}.
\newblock {The Tycho double star catalogue}.
\newblock {\em \aap}, 384:\penalty0 180--189, March 2002.

\bibitem[{Fabrycky} and {Tremaine}(2007)]{Fabrycky2007}
D.~{Fabrycky} and S.~{Tremaine}.
\newblock {Shrinking binary and planetary orbits by Kozai cycles with tidal
  friction}.
\newblock {\em ArXiv e-prints}, 705, May 2007.

\bibitem[{Fekel}(1980)]{Fekel1980}
F.~C. {Fekel}.
\newblock {The close multiple system MU Orionis - Masses and the
  mass-luminosity relation}.
\newblock {\em \pasp}, 92:\penalty0 785--789, December 1980.

\bibitem[{Fekel} et~al.(2002){Fekel}, {Scarfe}, {Barlow}, {Hartkopf}, {Mason},
  and {McAlister}]{Fekel2002}
F.~C. {Fekel}, C.~D. {Scarfe}, D.~J. {Barlow}, W.~I. {Hartkopf}, B.~D. {Mason},
  and H.~A. {McAlister}.
\newblock {The Quadruple System {$\mu$} Orionis: Three-dimensional Orbit and
  Physical Parameters}.
\newblock {\em \aj}, 123:\penalty0 1723--1740, March 2002.

\bibitem[{Frost}(1906)]{Frost1906}
E.~B. {Frost}.
\newblock {Spectrographic observations. Four stars with variable radial
  velocities.}
\newblock {\em \apj}, 23:\penalty0 264--269, April 1906.

\bibitem[{Girardi} et~al.(2002){Girardi}, {Bertelli}, {Bressan}, {Chiosi},
  {Groenewegen}, {Marigo}, {Salasnich}, and {Weiss}]{Girardi2002}
L.~{Girardi}, G.~{Bertelli}, A.~{Bressan}, C.~{Chiosi}, M.~A.~T. {Groenewegen},
  P.~{Marigo}, B.~{Salasnich}, and A.~{Weiss}.
\newblock {Theoretical isochrones in several photometric systems. I.
  Johnson-Cousins-Glass, HST/WFPC2, HST/NICMOS, Washington, and ESO Imaging
  Survey filter sets}.
\newblock {\em \aap}, 391:\penalty0 195--212, August 2002.

\bibitem[{Heintz}(1996)]{Heintz1996}
W.~D. {Heintz}.
\newblock {A Study of Multiple-Star Systems}.
\newblock {\em \aj}, 111:\penalty0 408--441, January 1996.

\bibitem[{Horch} et~al.(2001){Horch}, {van Altena}, {Girard}, {Franz}, {L{\'
  o}pez}, and {Timothy}]{Hor2001b}
E.~{Horch}, W.~F. {van Altena}, T.~M. {Girard}, O.~G. {Franz}, C.~E. {L{\'
  o}pez}, and J.~G. {Timothy}.
\newblock {Speckle Interferometry of Southern Double Stars. II. Measures from
  the CASLEO 2.15 Meter Telescope, 1995-1996}.
\newblock {\em \aj}, 121:\penalty0 1597--1606, March 2001.

\bibitem[{Horch} et~al.(2002){Horch}, {Robinson}, {Meyer}, {van Altena},
  {Ninkov}, and {Piterman}]{hor02}
E.~P. {Horch}, S.~E. {Robinson}, R.~D. {Meyer}, W.~F. {van Altena},
  Z.~{Ninkov}, and A.~{Piterman}.
\newblock {Speckle Observations of Binary Stars with the WIYN Telescope. II.
  Relative Astrometry Measures during 1998-2000}.
\newblock {\em \aj}, 123:\penalty0 3442--3459, June 2002.

\bibitem[{Hummel} et~al.(2003){Hummel}, {Benson}, {Hutter}, {Johnston},
  {Mozurkewich}, {Armstrong}, {Hindsley}, {Gilbreath}, {Rickard}, and
  {White}]{hum03}
C.~A. {Hummel}, J.~A. {Benson}, D.~J. {Hutter}, K.~J. {Johnston},
  D.~{Mozurkewich}, J.~T. {Armstrong}, R.~B. {Hindsley}, G.~C. {Gilbreath},
  L.~J. {Rickard}, and N.~M. {White}.
\newblock {First Observations with a Co-phased Six-Station Optical
  Long-Baseline Array: Application to the Triple Star {$\eta$} Virginis}.
\newblock {\em \aj}, 125:\penalty0 2630--2644, May 2003.

\bibitem[{Kozai}(1962)]{Kozai1962}
Y.~{Kozai}.
\newblock {Secular perturbations of asteroids with high inclination and
  eccentricity}.
\newblock {\em \aj}, 67:\penalty0 591--598, November 1962.

\bibitem[{Lane} and {Colavita}(2003)]{lc03}
B.~F. {Lane} and M.~M. {Colavita}.
\newblock {Phase-referenced Stellar Interferometry at the Palomar Testbed
  Interferometer}.
\newblock {\em \aj}, 125:\penalty0 1623--1628, March 2003.

\bibitem[{Lane} and {Muterspaugh}(2004)]{LaneMute2004a}
B.~F. {Lane} and M.~W. {Muterspaugh}.
\newblock {Differential Astrometry of Subarcsecond Scale Binaries at the
  Palomar Testbed Interferometer}.
\newblock {\em \apj}, 601:\penalty0 1129--1135, February 2004.

\bibitem[{Lane} et~al.(2007){Lane}, {Muterspaugh}, {Fekel}, {Williamson},
  {Browne}, {Konacki}, {Burke}, {Colavita}, {Kulkarni}, and
  {Shao}]{lane88Tau2007_draft}
B.~F. {Lane}, M.~W. {Muterspaugh}, F.~C. {Fekel}, M.~{Williamson}, S.~{Browne},
  M.~{Konacki}, B.~F. {Burke}, M.~M. {Colavita}, S.~R. {Kulkarni}, and
  M.~{Shao}.
\newblock The orbits of the quadruple star system 88 tau a from phases
  differential astrometry and radial velocity.
\newblock Submitted to {\apj}., 2007.

\bibitem[{Lestrade} et~al.(1993){Lestrade}, {Phillips}, {Hodges}, and
  {Preston}]{les93}
J.~{Lestrade}, R.~B. {Phillips}, M.~W. {Hodges}, and R.~A. {Preston}.
\newblock {VLBI astrometric identification of the radio emitting region in
  Algol and determination of the orientation of the close binary}.
\newblock {\em \apj}, 410:\penalty0 808--814, June 1993.

\bibitem[{Muterspaugh} et~al.(2005){Muterspaugh}, {Lane}, {Konacki}, {Burke},
  {Colavita}, {Kulkarni}, and {Shao}]{Mut05_delequ}
M.~W. {Muterspaugh}, B.~F. {Lane}, M.~{Konacki}, B.~F. {Burke}, M.~M.
  {Colavita}, S.~R. {Kulkarni}, and M.~{Shao}.
\newblock {PHASES High-Precision Differential Astrometry of {$\delta$}
  Equulei}.
\newblock {\em \aj}, 130:\penalty0 2866--2875, December 2005.

\bibitem[{Muterspaugh} et~al.(2006{\natexlab{a}}){Muterspaugh}, {Lane},
  {Konacki}, {Burke}, {Colavita}, {Kulkarni}, and {Shao}]{Mut06_v819her}
M.~W. {Muterspaugh}, B.~F. {Lane}, M.~{Konacki}, B.~F. {Burke}, M.~M.
  {Colavita}, S.~R. {Kulkarni}, and M.~{Shao}.
\newblock {PHASES differential astrometry and the mutual inclination of the
  V819 Herculis triple star system}.
\newblock {\em \aap}, 446:\penalty0 723--732, February 2006{\natexlab{a}}.

\bibitem[{Muterspaugh} et~al.(2006{\natexlab{b}}){Muterspaugh}, {Lane},
  {Konacki}, {Wiktorowicz}, {Burke}, {Colavita}, {Kulkarni}, and
  {Shao}]{Mut06_kappeg}
M.~W. {Muterspaugh}, B.~F. {Lane}, M.~{Konacki}, S.~{Wiktorowicz}, B.~F.
  {Burke}, M.~M. {Colavita}, S.~R. {Kulkarni}, and M.~{Shao}.
\newblock {PHASES Differential Astrometry and Iodine Cell Radial Velocities of
  the {$\kappa$} Pegasi Triple Star System}.
\newblock {\em \apj}, 636:\penalty0 1020--1032, January 2006{\natexlab{b}}.

\bibitem[{Muterspaugh} et~al.(2006{\natexlab{c}}){Muterspaugh}, {Lane},
  {Kulkarni}, {Burke}, {Colavita}, and {Shao}]{Mute06Limits}
M.~W. {Muterspaugh}, B.~F. {Lane}, S.~R. {Kulkarni}, B.~F. {Burke}, M.~M.
  {Colavita}, and M.~{Shao}.
\newblock {Limits to Tertiary Astrometric Companions in Binary Systems}.
\newblock {\em \apj}, 653:\penalty0 1469--1479, December 2006{\natexlab{c}}.

\bibitem[{Orlov} and {Petrova}(2000)]{orlov2000}
V.~V. {Orlov} and A.~V. {Petrova}.
\newblock {Dynamical Stability of Triple Stars}.
\newblock {\em Astronomy Letters}, 26:\penalty0 250--260, April 2000.

\bibitem[{Orlov} and {Zhuchkov}(2005)]{orlov2005}
V.~V. {Orlov} and R.~Y. {Zhuchkov}.
\newblock {Analysis of the Dynamic Stability of Selected Multiple Stars with
  Weak Hierarchy}.
\newblock {\em Astronomy Reports}, 49:\penalty0 201--216, March 2005.

\bibitem[{Pan} et~al.(1993){Pan}, {Shao}, and {Colavita}]{PanAlgol}
X.~{Pan}, M.~{Shao}, and M.~M. {Colavita}.
\newblock {High angular resolution measurements of Algol}.
\newblock {\em \apjl}, 413:\penalty0 L129--L131, August 1993.

\bibitem[{Perryman} et~al.(1997){Perryman}, {Lindegren}, {Kovalevsky}, {Hoeg},
  {Bastian}, {Bernacca}, {Cr{\' e}z{\' e}}, {Donati}, {Grenon}, {van Leeuwen},
  {van der Marel}, {Mignard}, {Murray}, {Le Poole}, {Schrijver}, {Turon},
  {Arenou}, {Froeschl{\' e}}, and {Petersen}]{hipcat}
M.~A.~C. {Perryman}, L.~{Lindegren}, J.~{Kovalevsky}, E.~{Hoeg}, U.~{Bastian},
  P.~L. {Bernacca}, M.~{Cr{\' e}z{\' e}}, F.~{Donati}, M.~{Grenon}, F.~{van
  Leeuwen}, H.~{van der Marel}, F.~{Mignard}, C.~A. {Murray}, R.~S. {Le Poole},
  H.~{Schrijver}, C.~{Turon}, F.~{Arenou}, M.~{Froeschl{\' e}}, and C.~S.
  {Petersen}.
\newblock {The HIPPARCOS Catalogue}.
\newblock {\em \aap}, 323:\penalty0 L49--L52, July 1997.

\bibitem[{Scardia} et~al.(2007){Scardia}, {Prieur}, {Pansecchi}, {Argyle},
  {Basso}, {Sala}, {Ghigo}, {Koechlin}, and {Aristidi}]{Sca2007a}
M.~{Scardia}, J.-L. {Prieur}, L.~{Pansecchi}, R.~W. {Argyle}, S.~{Basso},
  M.~{Sala}, M.~{Ghigo}, L.~{Koechlin}, and E.~{Aristidi}.
\newblock {Speckle observations with PISCO in Merate - III. Astrometric
  measurements of visual binaries in 2005 and scale calibration with a grating
  mask}.
\newblock {\em \mnras}, 374:\penalty0 965--978, January 2007.

\bibitem[{Skrutskie} et~al.(2006){Skrutskie}, {Cutri}, {Stiening}, {Weinberg},
  {Schneider}, {Carpenter}, {Beichman}, {Capps}, {Chester}, {Elias}, {Huchra},
  {Liebert}, {Lonsdale}, {Monet}, {Price}, {Seitzer}, {Jarrett}, {Kirkpatrick},
  {Gizis}, {Howard}, {Evans}, {Fowler}, {Fullmer}, {Hurt}, {Light}, {Kopan},
  {Marsh}, {McCallon}, {Tam}, {Van Dyk}, and {Wheelock}]{2MASS}
M.~F. {Skrutskie}, R.~M. {Cutri}, R.~{Stiening}, M.~D. {Weinberg},
  S.~{Schneider}, J.~M. {Carpenter}, C.~{Beichman}, R.~{Capps}, T.~{Chester},
  J.~{Elias}, J.~{Huchra}, J.~{Liebert}, C.~{Lonsdale}, D.~G. {Monet},
  S.~{Price}, P.~{Seitzer}, T.~{Jarrett}, J.~D. {Kirkpatrick}, J.~E. {Gizis},
  E.~{Howard}, T.~{Evans}, J.~{Fowler}, L.~{Fullmer}, R.~{Hurt}, R.~{Light},
  E.~L. {Kopan}, K.~A. {Marsh}, H.~L. {McCallon}, R.~{Tam}, S.~{Van Dyk}, and
  S.~{Wheelock}.
\newblock {The Two Micron All Sky Survey (2MASS)}.
\newblock {\em \aj}, 131:\penalty0 1163--1183, February 2006.

\bibitem[{Smith}(1983)]{MassLuminosity}
R.~C. {Smith}.
\newblock {An empirical stellar mass-luminosity relationship}.
\newblock {\em The Observatory}, 103:\penalty0 29--31, February 1983.

\bibitem[S{\" o}derhjelm(1999)]{Soder1999}
S.~S{\" o}derhjelm.
\newblock Visual binary orbits and masses post Hipparcos.
\newblock {\em A\&A}, 341:\penalty0 121--140, 1999.

\bibitem[{Sterzik} and {Tokovinin}(2002)]{Sterzik2002}
M.~F. {Sterzik} and A.~A. {Tokovinin}.
\newblock {Relative orientation of orbits in triple stars}.
\newblock {\em \aap}, 384:\penalty0 1030--1037, March 2002.

\bibitem[{van Hamme} et~al.(1994){van Hamme}, {Hall}, {Hargrove}, {Henry},
  {Wasson}, {Barkslade}, {Chang}, {Fried}, {Green}, {Lines}, {Lines},
  {Nielsen}, {Powell}, {Reisenweber}, {Rogers}, {Shervais}, and
  {Tatum}]{vanHamme1994}
W.~V. {van Hamme}, D.~S. {Hall}, A.~W. {Hargrove}, G.~W. {Henry}, R.~{Wasson},
  W.~S. {Barkslade}, S.~{Chang}, R.~E. {Fried}, C.~L. {Green}, H.~C. {Lines},
  R.~D. {Lines}, P.~{Nielsen}, H.~D. {Powell}, R.~C. {Reisenweber}, C.~W.
  {Rogers}, S.~{Shervais}, and R.~{Tatum}.
\newblock {The two variables in the triple system HR 6469 = V819 Her: One
  eclipsing, one spotted}.
\newblock {\em \aj}, 107:\penalty0 1521--1528, April 1994.

\end{thebibliography}
\bibliographystyle{plainnat}

\end{document}